\documentclass[doublespacing]{elsart}
\usepackage{threeparttable}

\usepackage{lineno}
\usepackage{natbib}
\usepackage{epsfig}

\usepackage{amssymb}

\begin{document}
\begin{frontmatter}
\title{Mesospheric vertical thermal structure and winds on Venus from HHSMT
CO spectral-line Observations}

\author[]{Miriam Rengel\corauthref{cor}} \corauth[cor]{Corresponding
author.},
\ead{rengel@mps.mpg.de}
\ead[url]{http://www.mps.mpg.de/homes/rengel/}
\author[]{Paul Hartogh},
\author[]{Christopher Jarchow}
\address{Max-Planck-Institut f\"ur Sonnensystemforschung,
                  Max-Planck-Strasse 2, \\37191, Katlenburg-Lindau, Germany}

\begin{abstract}

We report vertical thermal structure and wind velocities in the
Venusian mesosphere retrieved from carbon monoxide ($^{12}$CO J =
2–-1 and $^{13}$CO J = 2–-1) spectral line observations obtained
with the Heinrich Hertz Submillimeter Telescope (HHSMT). We observed
the mesosphere of Venus from two days after the second Messenger flyby
of Venus (on June 5 2007 at 23:10 UTC) during five days. Day-to-day and
day-to-night temperature variations and short-term fluctuations of
the mesospheric zonal flow were evident in our data. The extensive
layer of warm air detected recently by SPICAV at 90 -- to 100 km
altitude is also detected in the temperature profiles reported here.

 These data were part of a coordinated ground-based  Venus observational
campaign in support of the ESA Venus Express mission. Furthermore,
this study attempts to cross-calibrate space– and ground-based
observations, to constrain radiative transfer and retrieval
algorithms for planetary atmospheres, and to contribute to a more
thorough understanding of the global patterns of circulation of the
Venusian atmosphere.

\end{abstract}
\begin{keyword}
Venus \sep Middle Atmosphere


\end{keyword}

\end{frontmatter}
\begin{linenumbers}
\section{Introduction}\label{intro}

Understanding the Venusian atmospheric dynamics is one of the
fundamental problems in planetary sciences. Its study is of great
value not only to learn about Venus itself, but also to interpret
observations of extrasolar terrestrial planets, and among other examples, to understand
their range of habitability. Two international missions
recently joined efforts and carried out multi-point observations of
the Venusian atmosphere on 6 June for several hours: NASA's
MESSENGER spacecraft swung by Venus for a second time on 5 June 2007
at 23:10 UTC on its way to Mercury, and ESA's Venus Express is
orbiting around Venus since 11 April 2006. Among the space-based
observations, a world-wide coordinated Earth-based Venus
observational campaign from 23 May 2007 to June 9 (and later) was
initiated to remotely observe the Venusian
atmosphere\footnote{http://sci.esa.int/science-e/www/object/index.cfm?fobjectid=41012}.
Because Venus was close to its maximum eastern elongation during
June 2007, Venus was in a favourable position for ground-based
observations of both its day- and nightside.

The Venusian atmosphere is conventionally divided into three
regions: the troposphere (below 70 km), the mesosphere (70 - 120
km), and the thermosphere (above 120 km). Studying the Venus'
mesosphere dynamics is of special interest because this transition
region is characterized by the combination of two different wind
regimes (a retrograde super-rotation of the lower atmosphere, and a
sub-solar to anti-solar flow pattern, SS-AS, of the upper
atmosphere), and affects both the chemical stability and the thermal
structure of the entire atmosphere \citep{c03}. The principal
feature of Venus' atmospheric general circulation is the zonal
super-rotation with typical wind velocities of 60--120 m s$^{-1}$ at
the cloud top and nearly zero on the ground. An important (best
measured) tracer of the atmospheric motion of Venus is carbon monoxide (CO)
\citep{k76,w81}.
Sub-millimetre spectral line observations of CO and its isotopes
play an important role in the investigation of the poorly
constrained Venus mesosphere (it is the only technique to provide
direct wind measurements in the mesosphere): they are used to
retrieve vertical profiles of CO, temperature, and winds at the
mesospheric altitudes (e.g., \cite{cm91,l94,c03,r08}).

Data capabilities and partial preliminary results from the submillimetre
observations of the CO lines carried out with the Heinrich Hertz
Submillimeter Telescope (HHSMT) during  June 2007 on the mesosphere
of Venus as a part of the ground-based observing campaign were
reported in Rengel et al. (2008).

In this paper, we present a complete retrieval analysis of the  data
obtained, and report of mesospheric parameters such as wind,
vertical thermal structure, and CO abundance derived from the
$^{12}$CO J = 2--1 and $^{13}$CO J = 2--1 lines by the use of a
radiative transfer model and retrieval algorithm. Section \ref{obs}
summarizes the observations and describes the data reduction. The
results are presented in Section\,3. Spacial/temporal changes in the
mesospheric thermal structure and winds are discussed in Section\,4.

\section{Observations and data reduction}\label{obs}

CO observations on the disk of Venus were carried out at the HHSMT,
operated and owned by the Arizona Radio Observatory (ARO).
The telescope is located at an elevation of 3178 m on Mount Graham,
Arizona, and consists of a 10-m diameter primary with a nutating
secondary. Our 8, 9, 10, 14 and 15 June from 18:30 to 0:30 UT
observations employed the 345 Superconductor-Insulator
Superconductor (SIS) and the 2mmJT/1.3mmJT ALMA Sideband
Separating\footnote{developed as part of the ALMA project, this
system is the first of this kind to incorporate the latest SIS mixer
technology: image-separating mixers. Here, the image separating
system operates truly separating image noise and signal. It uses an
old 1.3mm and 2mm quasioptical JT Dewar and cross--grid to separate
the two orthogonal linear polarizations.} receivers, operating
respectively at 320--375 and 210--279 GHz to observe the $^{12}$CO J
= 2–-1 (at a frequency of 230.54 GHz), $^{12}$CO J = 3–-2 (at 345.79
GHz), and $^{13}$CO J = 2--1 (at 220.398 GHz) rotational
transitions. The 345\,SIS double sideband receiver was used with the
signal frequency being placed once in the lower sideband (LSB) and
another time in the upper sideband (USB), and the 2mmJT/1.3mmJT one
only in LSB.
We used simultaneously seven different backends: two 1 MHz Forbes
filterbanks (FFBA and FFBB), two 970 MHz wide
Acousto-Optical-Spectrometers (AOSA and AOSB, with mean resolutions
of 934 kHz for AOSA and 913 kHz for AOSB), two FB2 (FB2A and FB2B)
filterbackends, and one 215 MHz CHIRP Transform spectrometer (CTS,
resolution of $\sim$40\,kHz) \citep{h90,v06}.

Observing conditions were generally good, although  on 10, 14
and 15 June it was partially cloudy. System temperatures were typically
1500–-2500 for the 345 GHz receiver and 200–-500\,K for the
2mmJT/1.3mmJT receiver. The observing mode was always dual beam
switching. Pointing was checked every 2--3 h. The
typical integration time per individual spectrum was around 4\,min.

The Venus disc had a diameter of  23.44$''$  at the beginning and
25.55$''$ at the end of our observational period, with the evening
terminator separating the dayside crescent (the fraction of
illumination for Venus was 49.95 and 45.68\%, as seen by observer).
Fig.\,1 in Rengel et al. (2008) shows a synthetic image of the
apparent disc of Venus that approximates the telescopic view of
Venus as seen from the Earth at 8 June and 18:30 UT.

\begin{figure}[H]
\begin{center}
\psfig{file=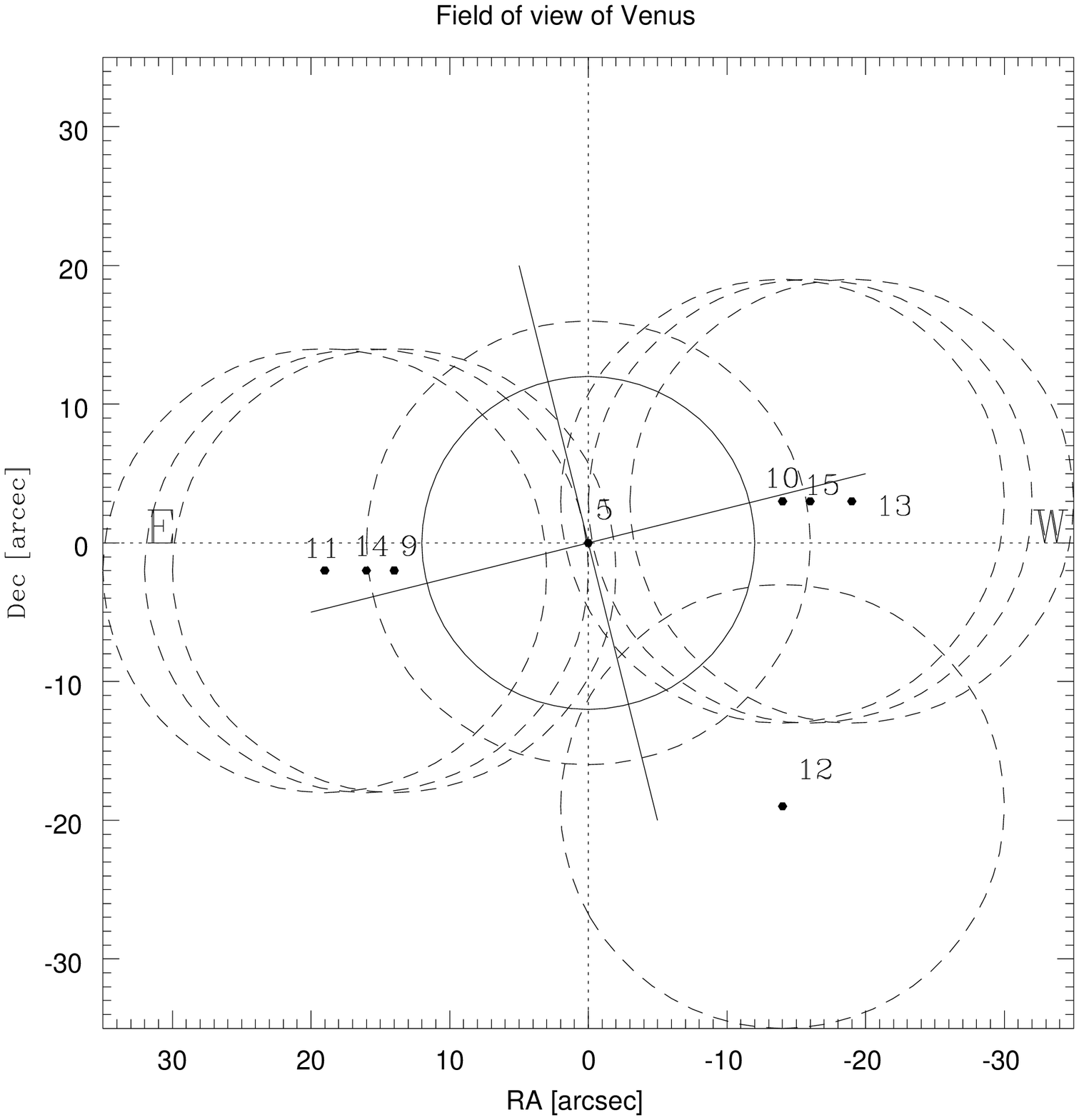,width=6cm} \psfig{file=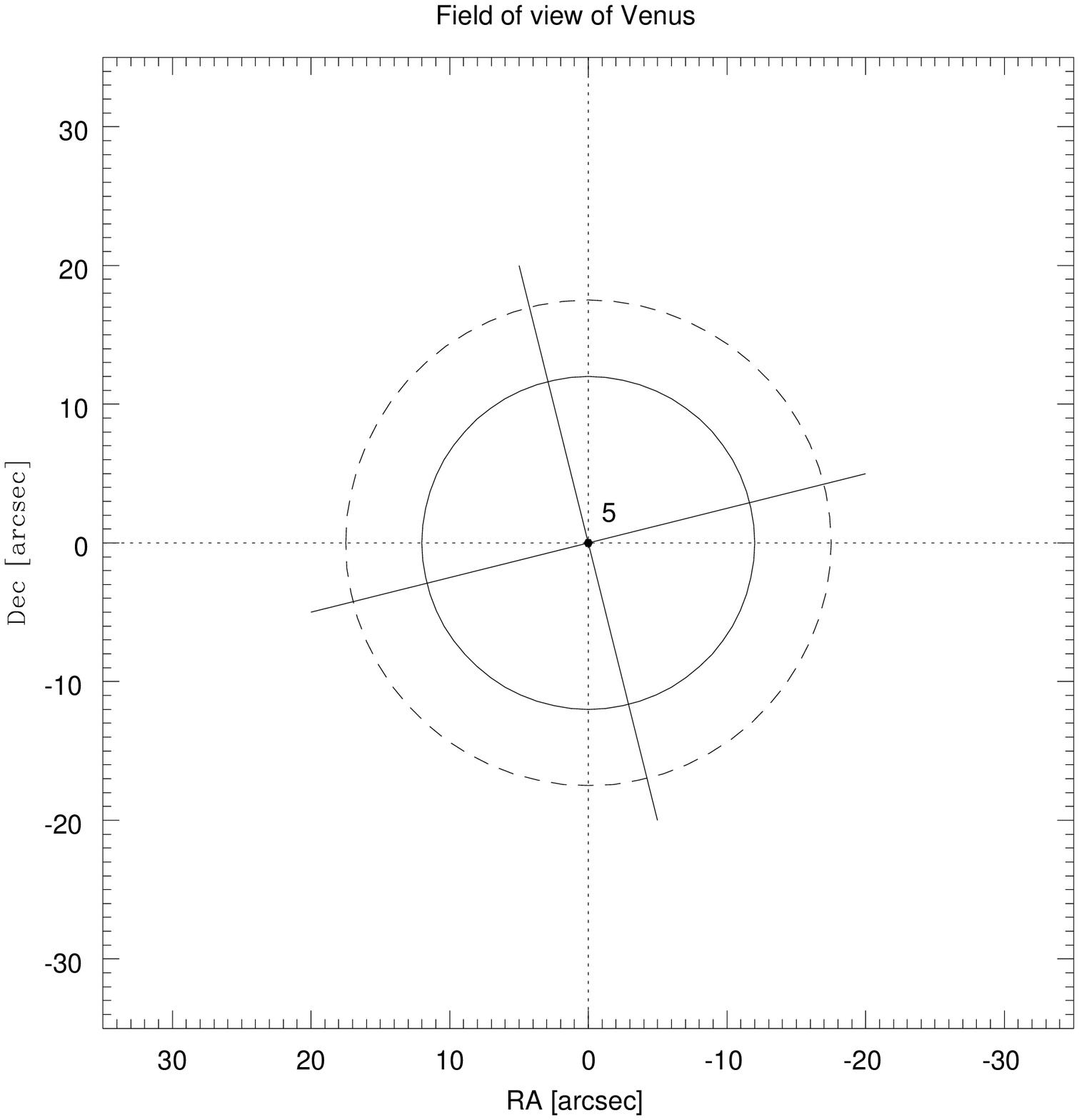,width=6cm}
\end{center}
\caption{} \label{pbeams}
\end{figure}

An overview of the complete data set can be found in Rengel et al.
(2008). We obtained measurements for $^{12}$CO J = 2–-1 at seven
positions (including centre, east and west limbs on the Venus disc)
and for $^{13}$CO J = 2–-1 at one (centre), see Fig.\,\ref{pbeams}.
The retrieval of the temperature profile and winds requires a clean
spectrum (spectrum with minor or almost zero baseline features), and
the CTS with a high dynamical range larger than 30\,dB is well
suited for our goals, here we concentrate only on the data taken
with the CTS. The $^{12}$CO J = 3–-2 line observations were
substantially noisier than $^{12}$CO J = 2–-1, so they are not
considered in our analysis here. Table\,1 gives a summary of our
$^{12}$CO J = 2–-1 and $^{13}$CO J = 2–-1 observations.

The initial reduction of each spectrum was performed using the CLASS
software package of the Grenoble Astrophysics
Group\footnote{http://www.iram.fr/IRAMFR/GILDAS}.

\section{Data analysis: retrieval of mesospheric
parameters (vertical thermal structure, CO abundance,  and winds)}


\subsection{Vertical thermal structure and CO profile}

In order to retrieve the temperature profile in the mesosphere, we
have applied a retrieval technique described by C. D. Rodgers as
optimal estimation \citep{r76}. We used a radiative transfer code
\citep{j95,j98,h04} which describes the physics of the radiative
transfer through the atmosphere, i.e., describes the transport and
balance of energy within the atmosphere, to calculate the synthetic
spectra which best fit the observed spectra. An a priori profile is
required as initial input for the optimal estimation technique. A
priori atmospheric temperature profiles, for day and nightsides,
were taken from the Venus international reference atmosphere (VIRA)
\citep{kliore}. For observations centred on the evening terminator,
a uniform profile at 172\,K at altitudes higher than 90\,km was
considered as approximation (and the VIRA profile at altitudes lower than 90\,km).
The a priori CO vertical profile employed is described in the paragraph below.
For the analysis here these profiles were automatically included in
the points distributed throughout the disc (Fig.\,2). Fig.\,2 shows
how we calculated the disc--averaged spectrum. For each black dot a
pencil beam spectrum (convolved with the telescope 2D Gaussian beam
pattern) has been calculated. The rings show the grouping of the
dots. Within a ring each dot has the same distance from the centre
of the planetary disc. The density of the dots has been chosen
according to the variation of the spectra: low density at the disc
centre and high density at the limb, because the spectra of the
single pencil beams vary only slowly at the centre, while due to the
exponential density decrease of the atmosphere with altitude the
limb area has to be resolved
much higher. 
We sub-divided the limb area into 16 rings to model accurately
the contribution of the limb region to the disc-averaged spectra.

Our atmospheric model consisted of 46 layers spanning the 30–-120 km
interval with a resolution of 2\,km. Upwelling continuum emission
from the $\sim$55\,km altitude region is modelled by incorporating
the collision induced absorption of CO$_2$ \citep{g97} and using the
VIRA temperature data at this altitude. Initially, the observed
spectra were modelled in units of percent absorption and then
converted to absolute brightness temperature by rescaling the
spectra in such a way that the continuum brightness temperature
equal to the VIRA temperature around 55\,km.

We derived the CO mixing ratio profile from the
simultaneous fitting of $^{12}$CO J = 2--1 and $^{13}$CO J = 2--1
HHSMT spectra (Fig.\,10, Obs. 35 and 36). Both observations
constrain CO abundances over different ranges of altitude since they
have a different opacity.
Two lines of the same molecule with strongly different altitude
dependent opacity allow retrieving simultaneously the vertical
profiles of temperature and its volume mixing ratio. Unfortunately
we have only two cases in which we observed two lines at the same
time. Obs.\,35, pointing only at the disc centre was the least noisy
one. We decided to retrieve the vertical profile of CO from this
observation and use it as a priori profile for all other
retrievals presented in this paper. We stress that simultaneous
temperature/CO retrievals are less reliable when only $^{12}$CO J =
2--1 data are available. The mentioned solely joint retrieval
employed as a priori a CO mixing ratio profile  purely increasing
exponentially with the altitude, a profile similar to the average of
the profiles of Figure 8\,A-B in Clancy and Muhleman (1991).
Note that we did not fix the CO profile with zero error
in the temperature from a single CO spectra line observation but
left error bars unequal zero on it, i.e., the error bars we got from
the CO retrieval. The reason not to fix the CO profile in the
retrieval (error bars zero) is that in reality the CO profile is
variable and each measurement, even with highest accuracy is just a
snapshot for the moment of the observation. In other words, for a
single $^{12}$CO line observation we give the fit algorithm the
freedom to not only fit the temperature profile, but to a small
extend also the CO profile. However, while for the CO profile
retrieval with two lines the correlation length for the CO profile
was only 10\,km (providing almost the same vertical resolution) in
the case of a single line retrieval the correlation length  has been
set to 40\,km. In other words, the fit algorithm can hardly modify
the (exponential) shape of the CO-profile, but almost just shift it
forth and back as a whole within the (relatively small) error bars.

The capability of the fitting algorithm to retrieve the atmospheric
temperature, CO-profile, and winds can be characterized by the
so-called weighting functions $K_x$ as defined by Rodgers 1990. Here
we briefly present the procedure we followed in order to calculate
$K_x$. Representing the atmosphere by a set of $N$ layers allows to
write the atmospheric profiles as a state vector $x$, where $x$ is a
vector of unknowns to be retrieved from the measurements $y$. $y$ is
a vector of measured quantities (i.e., the CO spectrum measured at
$m$ discrete frequencies). The so-called forward model $F(x)$
relates $x$ to $y$ as $y=F(x)$. This forward model contains the
complete physics necessary to calculate the absorption coefficient
as a function of temperature, mixing ratio, and spectral line
parameters, and in addition performs the radiation transfer through
the atmosphere. It is in general a non-linear function of the vector
$x$ of unknowns.

The weighting functions $K_x$ are now defined as  $K_x$=$\partial
F(x)/\partial x$. They describe the sensitivity of the measurement,
which means the observed CO spectrum, with respect to each single
parameter to be retrieved. Using $K_x$ allows to linearize the
forward model around a reference state $\bar{x}$:

\begin{equation}
y=F(\bar{x})+\frac{\partial F(\bar{x})}{\partial x}\,.\,(x-\bar{x}).
\end{equation}

For the calculation of $K_x$ here we used as reference state the a
priori profiles and normalized the resulting curves to unity as
maximum value. Because the measurement depends in a different way on
atmospheric temperature and CO mixing ratio, the weighting functions
have to be different for the temperature and mixing ratio. As seen
later, this is most strikingly visible at the altitude around
55\,km: here the collision induced absorption of CO$_2$ in Venus
atmosphere creates a sensitivity of the spectrum with respect to
temperature, independent of the CO mixing ratio. On the other hand,
the spectrum is not sensitive to the CO mixing ratio around this
altitude anymore, because the same CO$_2$ absorption lets the
atmosphere become optically thick below 55\,km. Finally, if the
weighting functions are zero in a certain altitude range this means
here the observation does not depend on the atmospheric parameters,
and consequently there parameters cannot be retrieved at this
altitude range.

Introducing the inverse model $I$, which relates the retrieved
atmospheric state $\hat{x}$ to $y$, $\hat{x}$ =$I(y)$, and combining
it with the forward model $y=F(x)$, allows to define a transfer
function $T$:

\begin{equation}
\hat{x}=I(F(x))=T(x).
\end{equation}

The averaging kernels $A$ are now defined as $A=\partial
T(x)/\partial x$ and describe the sensitivity of the retrieved
atmospheric parameters with respect to each single true parameter.
For each parameter, for example, a temperature at a certain altitude,
there is an averaging kernel function, showing that this temperature
depends on a change of the true temperature at all other altitudes.

Normalized temperature and CO-weighting functions for $^{12}$CO J =
2--1 (with separate panels for the nightside and dayside) and
$^{13}$CO J = 2--1 for several frequency offsets around each
transition are presented in Figs.\,\ref{wf1}--\ref{wf2}, and in
Fig.\,5, respectively. They show the vertical range and resolution
of temperature and of CO retrievals afforded here: temperature
sensitivity for $^{12}$CO J = 2--1  is primarily confined to the
55--110 km range, $^{12}$CO J = 2--1 sensitivity to 68--90 km range,
temperature sensitivity for $^{13}$CO J = 2--1 to 55--100 km range,
and $^{13}$CO J = 2--1 sensitivity to 70--95 km range. Because of
the input CO profile, the vertical distribution of the CO weighting
functions in Fig.\,\ref{wf2} are not greatly influenced by
day-to-night variation of CO abundances. However, the divergence of
day and night temperature profiles creates significant offsets in
day and night altitude scales above 90\,km altitude.

In Figs.\,6--11 we present the measured and fitted spectra,
and retrieved temperature profiles and CO distribution for the
observations in Table\,1 obtained with our technique (Figs.\,6--10
and Fig.\,11 (upper panels) for $^{12}$CO J = 2--1 and Fig.\,11
(lower panels) for $^{13}$CO J = 2--1 lines). Lower panels from the
spectra boxes in Figs.\,6--11 show difference between the observed
and fitted spectra, and averaging kernels for temperature, CO, and
winds, respectively. Note that the differences between the measured
and fitted spectra are so small that the they are hardly
distinguished when they are plotted simultaneously. For clarity we
show only each second averaging kernel curve.
Wind retrieval will be addressed in the next section. Error bars
represent the total error in the retrieval which comprises the
contributions of measurement errors, model errors,  and null-space
errors due to the inherent finite vertical resolution, as described
in \cite{r90}. These error terms appear as covariance matrices
(interpreted in terms of error patters) and are
equivalent to error bars. The contribution of a priori to  the
retrieval is described in terms of the eigenvectors of the
averaging kernel matrix. For the error bar calculations we followed
the recipe given in \cite{r90}. A discussion of the correlated
errors can be found in Sect.\,3.3.

The 15 June 2007 temperature profile from Obs.\,36, for example, is
compared with the ones obtained from SPICAV onboard Venus Express
\citep{n07}, Pioneer Venus (PV) descent probes \citep{s80}, to the
OIR sounding measurements \citep{st83}, and to the PV night probe
\citep{sek82} (Fig.\,\ref{compa}). We detected a temperature excess
at 90-- to 100\,km altitude which coincides with the extensive layer
of warm air at altitudes 90--120 km detected by SPICAV \citep{n07}.
However, its peak shows a smaller temperature excess than the one
detected by SPICAV. This particular layer has been interpreted as
the result of adiabatic heating during air subsidence. Because
adiabatic heating is a localized phenomenon, does the layer's
altitude increase with the latitude? We note that layer´s altitudes
of around 95, 97 and 100\,km are determined at latitudes of
4$^{\circ}$ S, 0$^{\circ}$, and 39$^{\circ}$ N, respectively
(\cite{n07}; this work). We stress this possibility, however,
three-point statistics cannot make a strong statement, additional
data at different latitudes are required.
Furthermore, in other altitudes the HHSMT profile compares favourably
to those returned by the previous measurements. Similar temperature
profiles were also observed \citep{l94,c03}. Moreover, it was
suggested that a 10--15\,K increase in the mesospheric temperatures
occur over 1--30--day periods, and large variations (20-40\,K) over as
yet undetermined timescales \citep{c03}. More data over longer
periods are necessary.

\subsection{Wind velocity retrieval}

Fig.\,13 shows the observed and fitted spectrum of $^{12}$CO J = 2--1 (Obs.\,27),
the residual (difference between observation and fit), and the retrieved vertical
temperature and CO profiles and respective kernels. The double peak in the centre
of the residuum (left lower panel) shows that the fit between the observed and synthetic
spectra  has deviations, meaning that some physical parameters are still being missing in the
model.
Obviously, this double peak is related to a frequency shift of the
fitted line relative to the observed one. This frequency shift is
caused by movements of the observed air mass along the line of
sight. In other words, we see the effect of winds causing Doppler
shifts in the centre of the observed spectrum. In order to get a
perfect fit an additional fit parameter has to be introduced, which
minimizes the residual by frequency shifting the spectrometer
channels around the line centre related to the wind speed as
function of altitude. In more detail, we retrieved the
beam-integrated wind velocities taking into account vertical
information. The model assumes an individual uniform zonal wind
speed within each atmospheric layer, and the layers together rotate
with individual angular velocities among a common axis, the
rotational axis of Venus. According to the component of the wind
velocity vector along the line of sight, each location on the Venus'
disc has its own Doppler shift: it is maximum at the limb on the
equator, and zero for all points of the Venus' disc falling onto the
rotational axis. Because the size of the main beam here is
comparable to the diameter of the Venus' disc for our observational
period, we integrated across the Venus' disc (see Fig. 2 and
explanation in the last chapter) to take into account the smoothly
varying Doppler shifts. Pointing exactly to the centre of the disc
just causes the spectral line slightly to broaden, because the
contributions from the western and eastern limb are Doppler-shifted
with respect to the line centre, but in opposite directions.
However, pointing to the eastern with respect to western limb causes
the contribution from this limb to dominate over the contribution
from the opposite limb (larger weight due to the antenna pattern)
and the resulting spectral line appears to be completely red, respectively.
blue shifted. This shift with respect to the nominal line centre is
sufficient to retrieve a wind velocity within each individual
atmospheric shell. As a priori profile we have chosen a wind speed
of 0 m\,s$^{-1}$ throughout the whole atmosphere, with a standard
deviation of 100 m\,s$^{-1}$. The information gained by the
observations will reduce the error bars in the altitude range the
observations are sensitive to. Both, the upper and the lower boundary
are related to spectral noise. If it exceeds a certain value, the
retrieval algorithm cannot relate Doppler shifts to wind speeds
smaller than 100 m\,s$^{-1}$. Concerning the wind profile vertical
resolution we have assumed a correlation between the layers
characterized by a correlation length of 4\,km (Rodgers 1990). This
value is similar to the one assumed for the temperature profile
resolution (for the CO mixing ratio we assumed higher values than
these ones in order to restrict the degrees of freedom of the
fitting). Wind-weighting functions for $^{12}$CO J = 2--1 line (day-
and nightsides) are presented in Fig.\,14, respectively. Wind
sensitivity is primarily confined to the 85-110\,km range. The
altitude resolution indicated by the averaging kernels is consistent
with the one indicated by the weighting functions. Both, the
weighting functions and the averaging kernels show a clear day/night
dependence of the altitude in which the winds are detected. Vice
versa, from the averaging kernels day and night observations can
easily be identified. Fig.\,6 shows wind retrievals for the Obs. 19,
20 and 21. Obs. 19 is pointing on the disc centre. The retrieved
wind speed is zero as expected. Obs. 20 is pointing on the night
side and provides a wind speed of nearly 300 m\,s$^{-1}$ at 110\,km
altitude. The non-zero residual can be interpreted in a way, that
for this observation the dayside emission is very weak and cannot
compensate the emission from the nightside, therefore one-half of
the double peak cannot be fitted.
The wind of Obs. 22 is comparable with that of Obs. 20.

All the wind measurements given here are deprojected ones.
Figs.\,6--11 present all retrieved wind velocities, from two
observing days (14 and 15 June), and Fig.\,\ref{wind2} retrieved
evident wind measurements as a function of beam position. In
Fig\,\ref{wind2}, when several observations were carried out at the
same beam position and day (Obs. 20, 22 and 24 at beam position 9,
Obs 21 and 29 for beam position 10, and Obs. 25 and 27 for beam
position 11), an averaged wind velocity is calculated (we assume
that the beam is positioned at the same right ascension for the
positions 9 and 14, and also for 10 and 15).
Strong winds are also reported by Lellouch et al. 2008 (this issue).



\subsection{Characterization of the correlated errors}

The optimal estimation method as described by Rodgers 1976 does not
only allow to retrieve a temperature, mixing ratio, and wind
profiles from the observed spectra of Venus, but in addition
automatically provides an error estimate for the derived quantities.
If each of the three profiles is represented by a set of $n$
discrete numbers corresponding to $n$ discrete layers representing
the atmosphere of Venus, then the error estimation is given in the
form of a covariance matrix $S$ with the dimension
3$n$\,$\times$3$n$. Its diagonal elements are the variances of the
profile values in each single layer and the square root of these
variances is shown in Figs.\,6--11 as error bars for the retrieved
profiles. However, if the off-diagonal elements of $S$ are not zero,
they indicate a correlation of these errors, for example, the
temperature profile in a certain layer with the temperature errors
within all the layers. In addition, the off-diagonal elements also
specify the correlation between the temperature, mixing ratio and
wind profile errors. In order to characterize these correlated
errors we followed the formal procedure described in Rodgers 1990
and performed as an example an eigenvector $\lambda$ and eigenvalue
$l$ analysis of $S$ for the Obs.\,34. This analysis provides the
so-called error patterns (orthogonal quantities
$e_i$=$\lambda_i^{1/2}\,l_i$) which show the correlation of the
errors for all retrieved quantities. There are in total 3$n$ error
patters and each single pattern consists of a part of $n$ numbers
belonging to the temperature, $n$ numbers to the mixing ratio, and
$n$ numbers to wind profiles. Because these three blocks have
different physical units for plotting purpose, a single pattern has
been split and plotted into three separate panels (left, centre and
right) in Figs.\,\ref{cerr1}--\ref{cerr2}. Looking through all the
error patterns suggests to sort them into three different groups:
error patterns with large amplitudes for only (1) the temperature
part, (2) the mixing ratio part, and (3) wind profile part. These
three groups are shown in Fig.\,\ref{cerr1}, upper, middle and lower
panels. The fact that such a separation is possible shows that the
errors of the three profiles are basically not correlated. The
oscillating nature of part of the error patters for the wind
profile, for example, indicates the strong correlation between the
wind profile errors in different layers. Looking at the error
patters with large amplitudes for the temperature reveals (1) the
amplitudes are only large for altitudes outside the 70--100\,km
altitude range, showing that the CO spectral line contains enough
information to significantly reduce the a priori profile error at
this altitude; (2) there exists one error pattern which behaves
exactly the opposite (see Fig.\,\ref{cerr1}): it has a large
amplitude within the 70--100\,km range, but a small one outside. At
the same time there is also a large error pattern amplitude for the
CO profile part, which means that only in this case there is a
strong correlation between the two errors. This single error pattern
describes most of the error of the temperature profile in that
altitude region, in which the observation is sensitive to
atmospheric temperature. A further discussion of error patterns due
to instrumental noise and null-space, and the effect of a priori
data can be also based in the procedure described in Rodgers 1990,
but this is beyond the scope of this paper.


\section{Discussion}
\subsection{Short-term changes in the mesospheric thermal structure and in winds}

We compare the day and night mesospheric temperature profiles from
the 14 and 15 June observing period in Fig.\,18. Direct detailed
shape comparison indicates day-to-night and day-to-day small
variations. For example, largest day-to-night variations of around
25\,K occur at 102\,km, day-to-day variations of around 15\,K occur
also at 85\,km, and night-to-night variations of around 15\,K occur
also at 102\,km. It must be stressed however that any variation
(spatial or temporal ones) is not differentiated here.
Furthermore, measurements acquired at longer observing periods over
the Venus' disc would define a pattern of variation (if any).

From Figs.\,6--11, a qualitative comparison of the wind measurements
suggests a day-to-day variability of the winds. For instance, for
beam position 9, the velocity increases from around 280 m\,s$^{-1}$
to around 355 m\,s$^{-1}$ from 14 to 15 June, and for beam position
10, from -68 to -280 m\,s$^{-1}$ from 14 to 15 June. From
West-to-East limb wind comparisons (at beam positions 10 and 13), we
detect negative weaker winds of -68$\pm$89 m\,s$^{-1}$ and of
-136$\pm$86 m\,s$^{-1}$, respectively. Furthermore, at beam
positions 9 and 11, we detect positive winds of 280$\pm$66
m\,s$^{-1}$ and of 205$\pm$76 m\,s$^{-1}$, respectively. Wind
measurements in Fig.\,\ref{wind2} are taken at the wind peak of
Figs.\,6--11.


\section{Conclusion}

We have carried out several HHSMT $^{12}$CO J = 2--1 and $^{13}$CO J
= 2--1  line observations on different beam positions on Venus disc
during June 2007 around the MESSENGER flyby of Venus and
observations from Venus Express mission.
From the spectra we
retrieved vertical profiles of temperature, CO distribution, and
wind velocities for the 2007 June mesosphere of Venus. Changes in
the thermal structure of the Venus mesosphere are detected:
day-to-night small temperature variations and short-term (day-to-day) on a
time scale as short as one day. This is consistent with the picture of
dramatic variability of the Venus mesosphere with changes in
temperature occurring on short scales (\cite{c05,s05}). Furthermore,
retrieved winds show variations of around 100 m\,s$^{-1}$ between
the winds on 14 June and those on 15 June.

HHSMT line observations of $^{12}$CO J = 2--1 and $^{13}$CO J = 2--1
and retrieved thermal profiles of the mesosphere of Venus (the
temperature peak detection at 90-100\,km) seems to support the new
finding of the extensive layer of warm air detected by SPICAV
onboard Venus Express.

Despite the success of the analysis presented here, some points need
further work. Wind measurements and fluctuations of the zonal flow
will be connected to possible CO distribution variability.

\section*{Acknowledgments}

We thank to the staff of the HHSMT for crucial support while
observing, and to J.-L Bertaux and to F. Montmessin for providing us
the SPICAV data parallel to publication. We also thank the two
manuscript referees for their criticism and valuable suggestions for
improvement of the initial submission. We acknowledge JPL's Horizons
online ephemeris generator for providing Venus's position during the
observations. This research has made use of NASA's Astrophysics Data
System.


\bibliographystyle{elsart-harv}
\bibliography{ref}

Table 1: Observation Parameters for the $^{12}$CO J =
2--1 line
\\
Fig. 1: Schematic of the beam locations on the Venus disc (for a 24
'' disc diameter) where $^{12}$CO J=2--1 and $^{12}$CO J=3--2 line
integrations are obtained (left and right panels, respectively).
Solid lines indicate the Venus'equator and central meridian. Dashed
circles indicate the approximate FWHM beam diameter.
\\
Fig. 2: Schematic representation of the Venus disc showing the
distribution of the points where a pencil beam spectrum has been
calculated. The rings show the grouping of the dots. Within a ring
each dot has the same distance from the centre of the Venus disc.
Limb area is not to scale with respect to the disc. Shade area
represents the disc corresponding to the solid body.
\\
Fig. 3: Dayside (left) and nightside (right) normalized atmospheric
temperature weighting functions as provided by $^{12}$CO J = 2--1
line. Each function corresponds to a particular frequency offset
from line centre (from 0 to 500 MHz offset).
\\
Fig. 4: Dayside (left) and nightside (right) normalized CO mixing
weighting functions as provided by $^{12}$CO J = 2--1 line. Each
function corresponds to a particular frequency offset from line
centre (from 0 to 500 MHz offset).
\\
Fig. 5: Temperature (left) and CO mixing weighting functions (right)
as provided by $^{13}$CO J = 2--1 line. Each function corresponds to
a particular frequency offset from line centre (from 0 to 500 MHz
offset).
\\
Fig. 6: Synthetic spectra solution for observations in Table\,1
(left upper panels), difference between the observed and fitted
spectra (left lower panels), retrieved temperature, CO distribution
and wind derived from the spectrum. The dash lines show the initial
profiles, and the horizontal lines the error bars.
\\
Fig. 7: continued
\\
Fig. 8: continued
\\
Fig. 9: continued
\\
Fig. 10: continued
\\
Fig. 11: continued
\\
Fig. 12: Temperature profile retrieval (from Obs.\,36), solid line,
compared to the profile from the stellar occultations with the
SPICAV onboard Venus Express \citep{n07}, PV descent probes
\citep{s80}, from the OIR sounding measurements \citep{st83}, and
from the PV night probe \citep{sek82}.The SPICAV measurements were
taken at latitude 39$^{\circ}$\,N for orbits 95, 96, and 98, and
latitude 4$^{\circ}$\,S for orbits 102-104. The Pioneer-Venus
derived VIRA reference profile for latitudes $<$30 are indicated by
the squares. The anomalously warm temperatures returned by the
Venera\,10 probe in 1975 are shown as stars symbols. The absolute
uncertainty for the temperatures derived here is $\pm$15\,K.
\\
Fig. 13: Observed and fitted spectrum of $^{12}$CO J = 2--1 (Obs.\,27, top left panel),
the residual (difference between observation and fit, bottom left panel), and the retrieved vertical
temperature and CO profiles (top centre and right panels) and respective kernels (low centre and right panels).
The double peak in the centre
of the residuum (left lower panel) is related to a frequency shift of the fitted line relative
to the observed one: effect of winds causing Doppler shifts in the centre of the observed spectrum.
\\
Fig. 14: Dayside (left) and nightside (right) wind weighting
functions as provided by $^{12}$CO J = 2--1 lines.
\\
Fig. 15: Retrieved wind velocity measurements for different beam
positions on the Venus' disc for two observing days.  Squares: 14 June, Triangles: 15 June. Retrieved velocity error bars are indicated.
E: East, W: West.
\\
Fig. 16: Error patterns belonging to the temperature, mixing ratio
and wind parts (upper, middle, and lower panels).
\\
Fig. 17: Error patterns for the Obs.\,34.
\\
Fig. 18: Dayside (left) and nightside (right) HHSMT temperature
profiles on 14 June and 15 June (Obs.\,29 and 34 for dayside with thin
solid and dot-short dash lines, and Obs. 22 and 33 for nightside
with thick solid and dot-short dash lines, respectively).
Measurements from SPICAV \citep{n07} for orbits 102--104 are shown
with dot-short, short dash-long, and dot-long dash lines,
respectively, and from the PV night probe \citep{s80} with
triangles.

\clearpage

\begin{table}
\caption{Observation Parameters}\
\begin{tiny}
\begin{threeparttable}
\begin{tabular}{crrccccc} \hline \hline
Beam          & d$\alpha$\tnote{a} & d$\delta$\tnote{a} & Obs. No. &  Scan No. &  Line  &  Date     &   Receiver/Sideband\\
\tiny Position      &           &           &          &           &        &  June 2007&                    \\
\hline
5  &   0    &  0   &  19 &  139--140&  $^{12}$CO J = 2–-1  &  14  & 2mm/1.3mm ALMA -- LSB\\
9  &   14   &  -2  &  20 &  141--146&  $^{12}$CO J = 2–-1  &  14  & 2mm/1.3mm ALMA -- LSB\\
10 &   -14  &  3   &  21 &  147--152&  $^{12}$CO J = 2–-1  &  14  & 2mm/1.3mm ALMA -- LSB\\
9  &   14   &  -2  &  22 &  153--156&  $^{12}$CO J = 2–-1  &  14.  & 2mm/1.3mm ALMA -- LSB\\
5  &   0    &  0   &  23 &  164--165&  $^{12}$CO J = 2–-1  &  14  & 2mm/1.3mm ALMA -- LSB\\
9  &   14   &  -2  &  24 &  166--167&  $^{12}$CO J = 2–-1  &  14  & 2mm/1.3mm ALMA -- LSB\\
11 &   19   &  -2  &  25 &  168--173&  $^{12}$CO J = 2–-1  &  14  & 2mm/1.3mm ALMA -- LSB\\
12 &   -14  &  -19 &  26 &  174--176&  $^{12}$CO J = 2–-1  &  14  & 2mm/1.3mm ALMA -- LSB\\
11 &  19    &  -2  &  27 &  177--182&  $^{12}$CO J = 2–-1  &  14  & 2mm/1.3mm ALMA -- LSB\\
13 &   -19  &  3   &  28 &  183--188&  $^{12}$CO J = 2–-1  &  14  & 2mm/1.3mm ALMA -- LSB\\
10 &   -14  &  3   &  29 &  189--194&  $^{12}$CO J = 2–-1  &  14  & 2mm/1.3mm ALMA -- LSB\\
5  &   0    &  0   &  30 &  195--196&  $^{12}$CO J = 2–-1  &  14  & 2mm/1.3mm ALMA -- LSB\\
5  &   0    &  0   &  31 &  199--202&  $^{13}$CO J = 2–-1  &  14  & 2mm/1.3mm ALMA -- LSB\\
5  &   0    &  0   &  32 &  208--217&  $^{12}$CO J = 2–-1  &  15  & 2mm/1.3mm ALMA -- LSB\\
14 &   16   &  -2  &  33 &  218--227&  $^{12}$CO J = 2–-1  &  15  & 2mm/1.3mm ALMA -- LSB\\
15 &   -16  &  +3  &  34 &  228--237&  $^{12}$CO J = 2–-1  &  15  & 2mm/1.3mm ALMA -- LSB\\
5  &   0    &  0   &  35 &  244--293&  $^{13}$CO J = 2–-1  &  15  & 2mm/1.3mm ALMA -- LSB\\
5  &   0    &  0   &  36 &  295--299&  $^{12}$CO J = 2–-1  &  15  & 2mm/1.3mm ALMA -- LSB\\
\hline
\end{tabular}
\begin{tablenotes}
       \item[a] d$\alpha$ and d$\delta$, right ascension and
declination, are the astronomical coordinates of a point on the
celestial sphere when using the equatorial coordinate system. The
earlier coordinate is the celestial equivalent of terrestrial
longitude, and the later one, to the latitude, projected onto the
celestial sphere.
     \end{tablenotes}
  \end{threeparttable}
\label{op}
\end{tiny}
\end{table}
\newcommand{\degree}{\ensuremath{^\circ}}

\normalsize

\begin{figure}
\begin{center}
\psfig{file=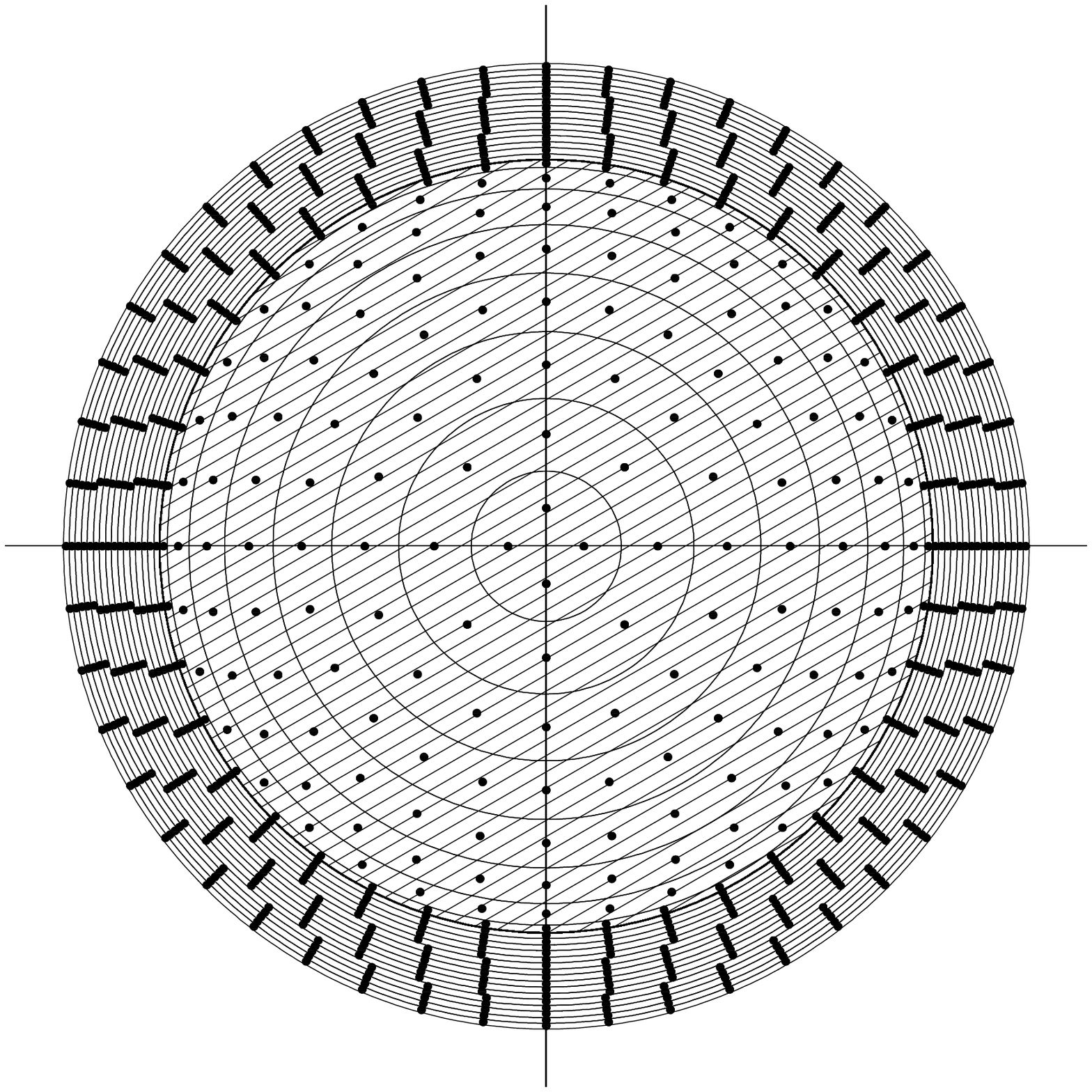,width=11cm}
\end{center}
\caption{} \label{dp}
\end{figure}

\begin{figure}[H]
\begin{center}
\psfig{file=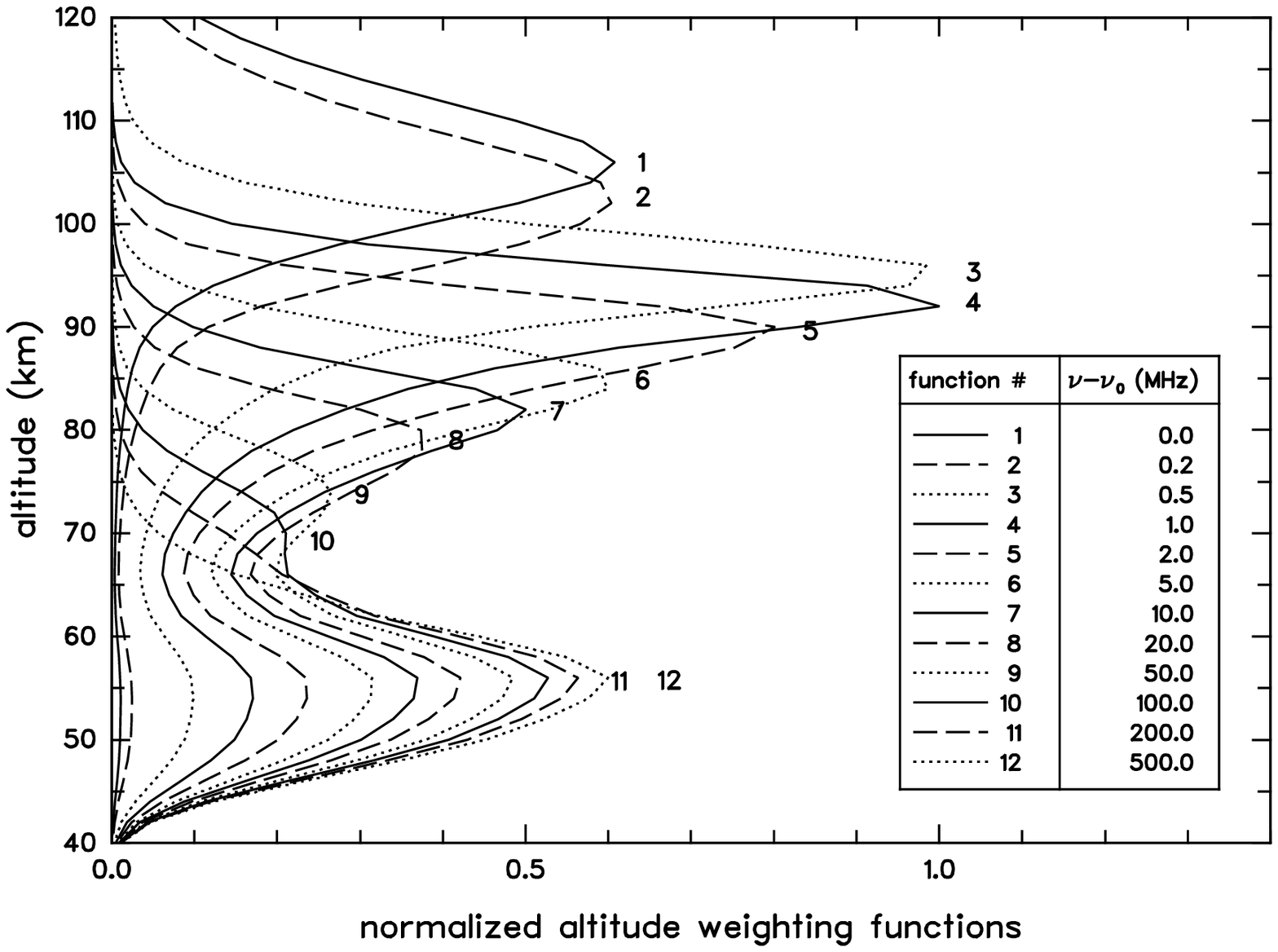,width=6cm}
\psfig{file=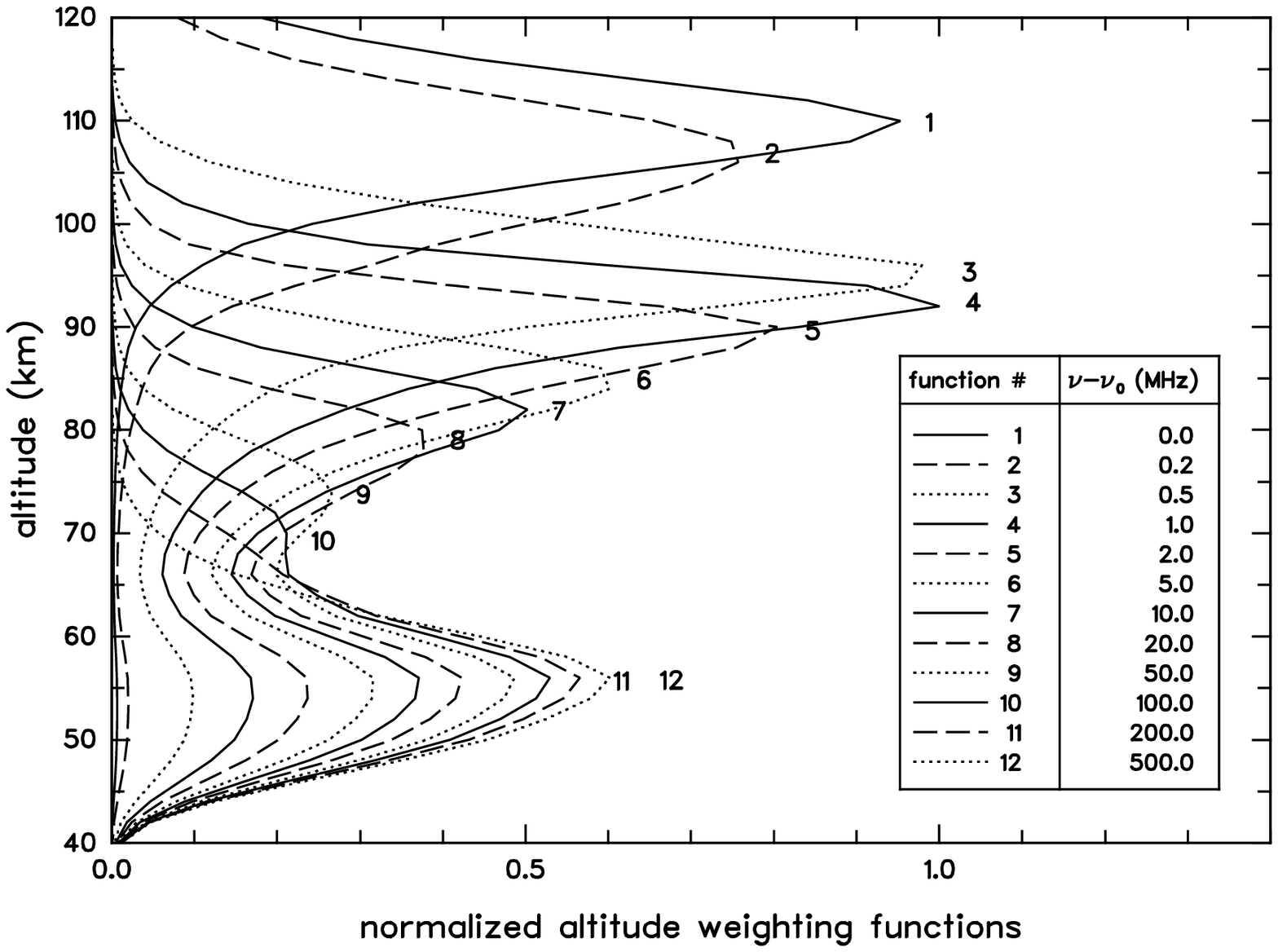,width=6cm}
\end{center}
\caption{} \label{wf1}
\end{figure}

\begin{figure}[H]
\begin{center}
\psfig{file=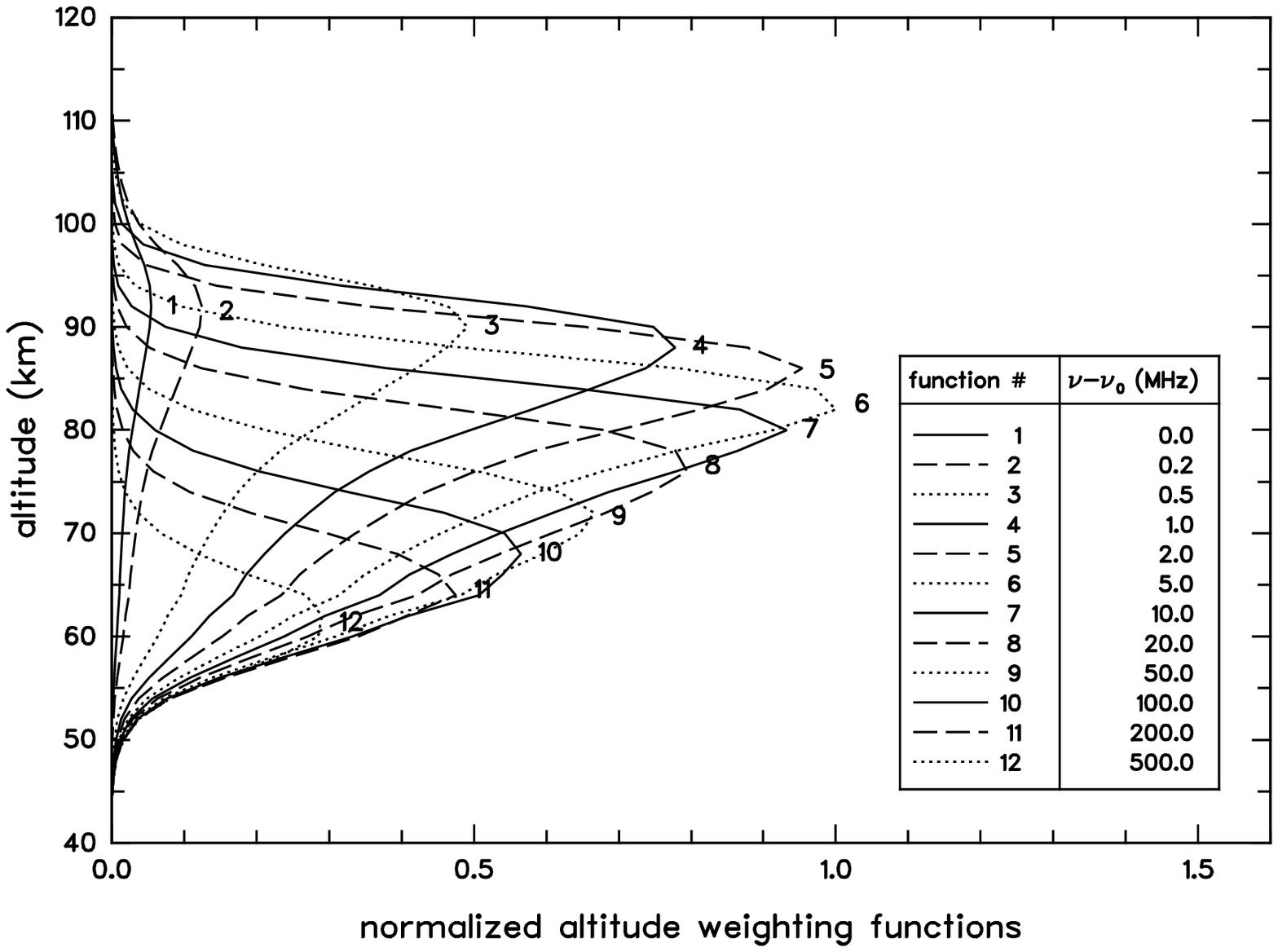,width=6cm}
\psfig{file=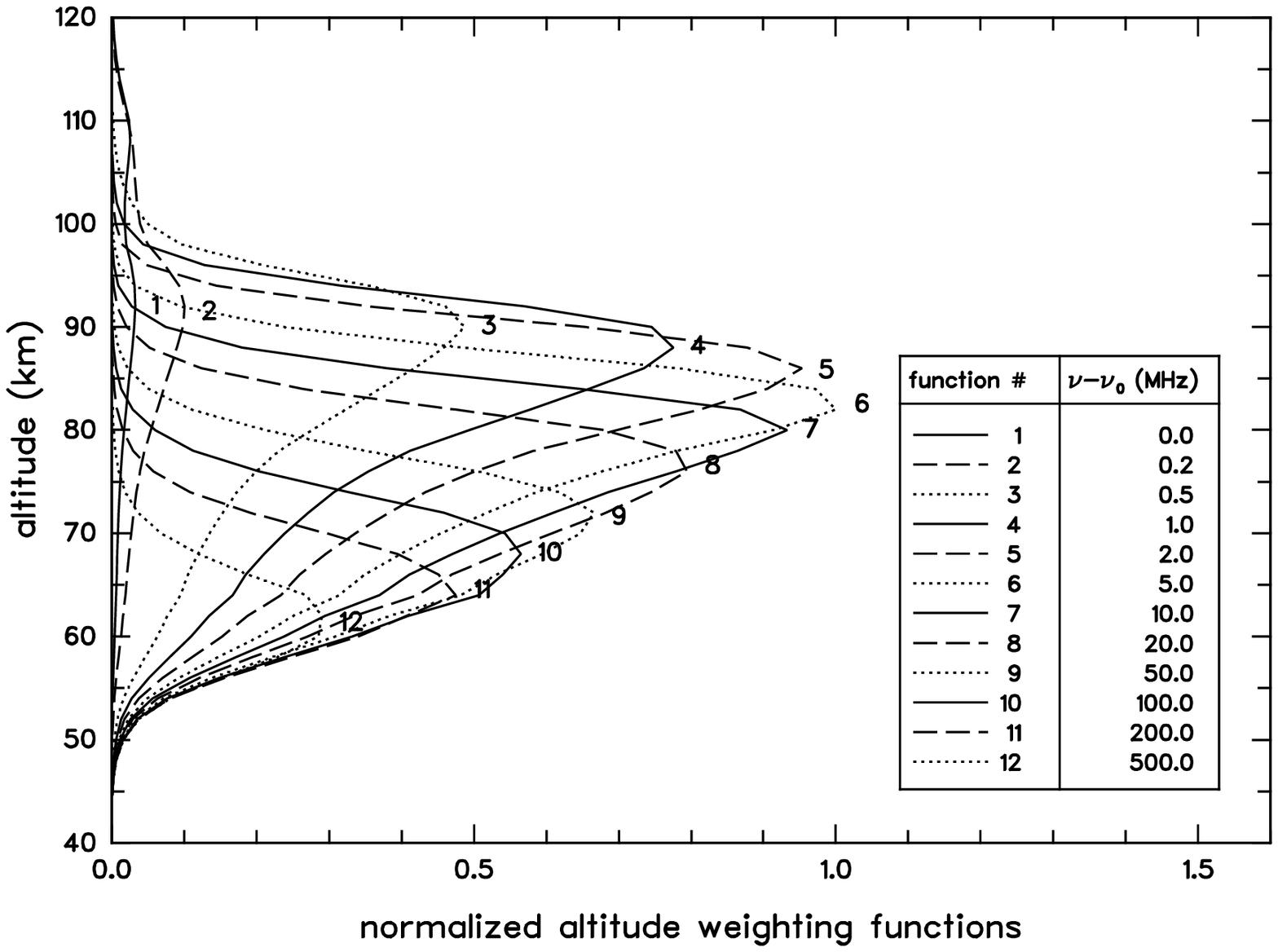,width=6cm}
\end{center}
\caption{} \label{wf2}
\end{figure}

\begin{figure}[H]
\begin{center}
\psfig{file=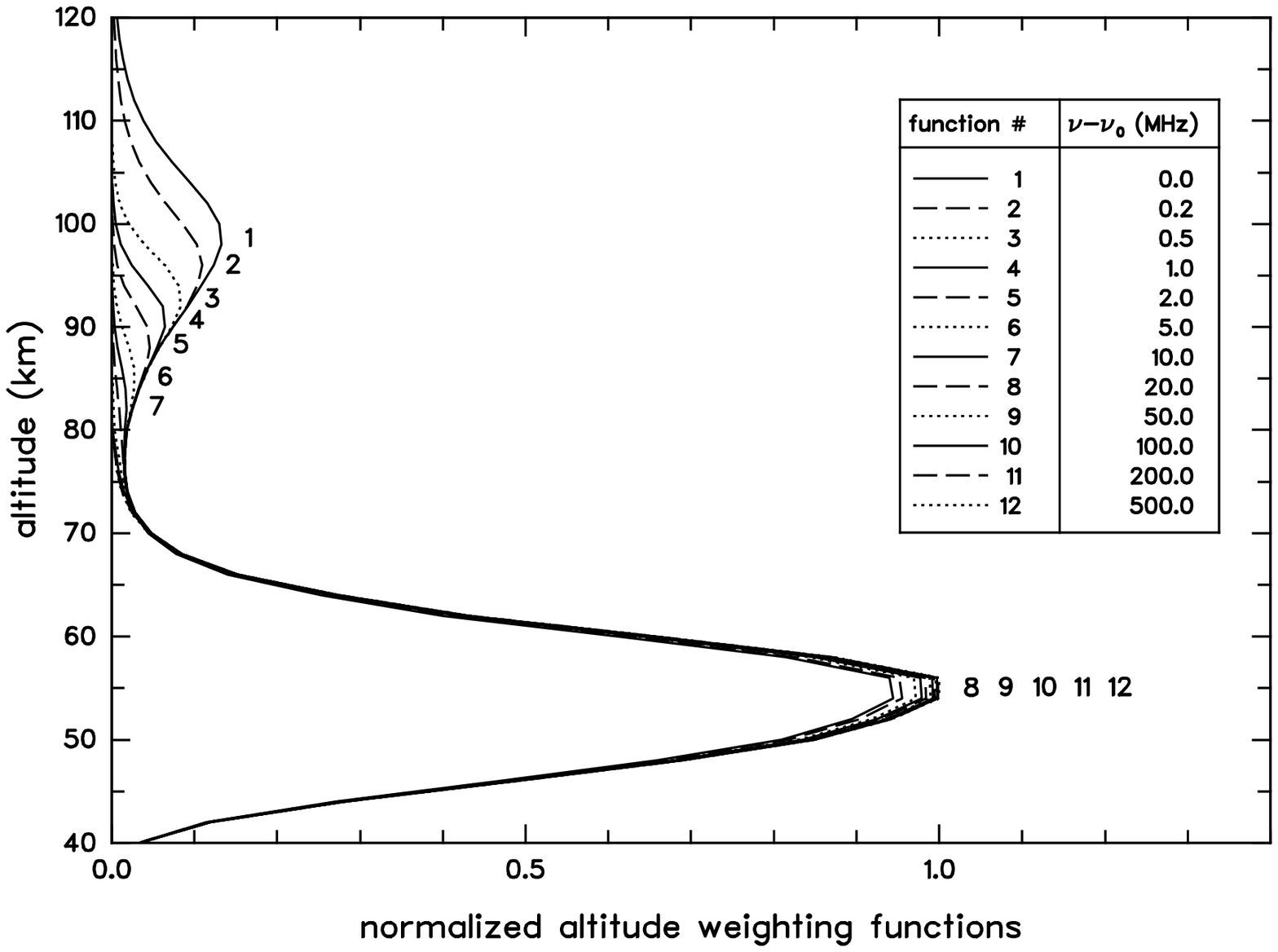,width=6cm}
\psfig{file=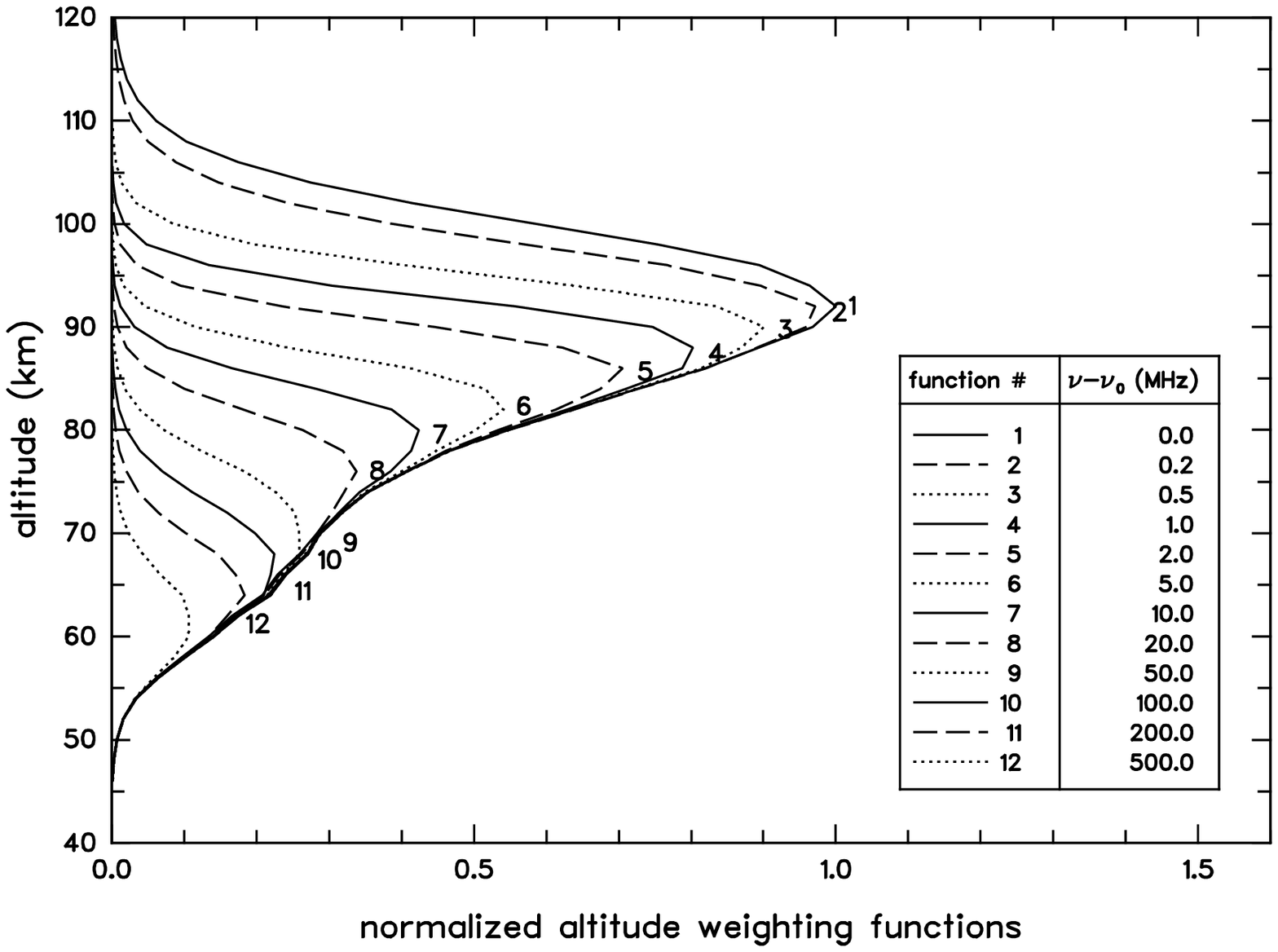,width=6cm}
\end{center}
\caption{} \label{wf3}
\end{figure}

\begin{figure}[H]
\begin{center}
\psfig{file=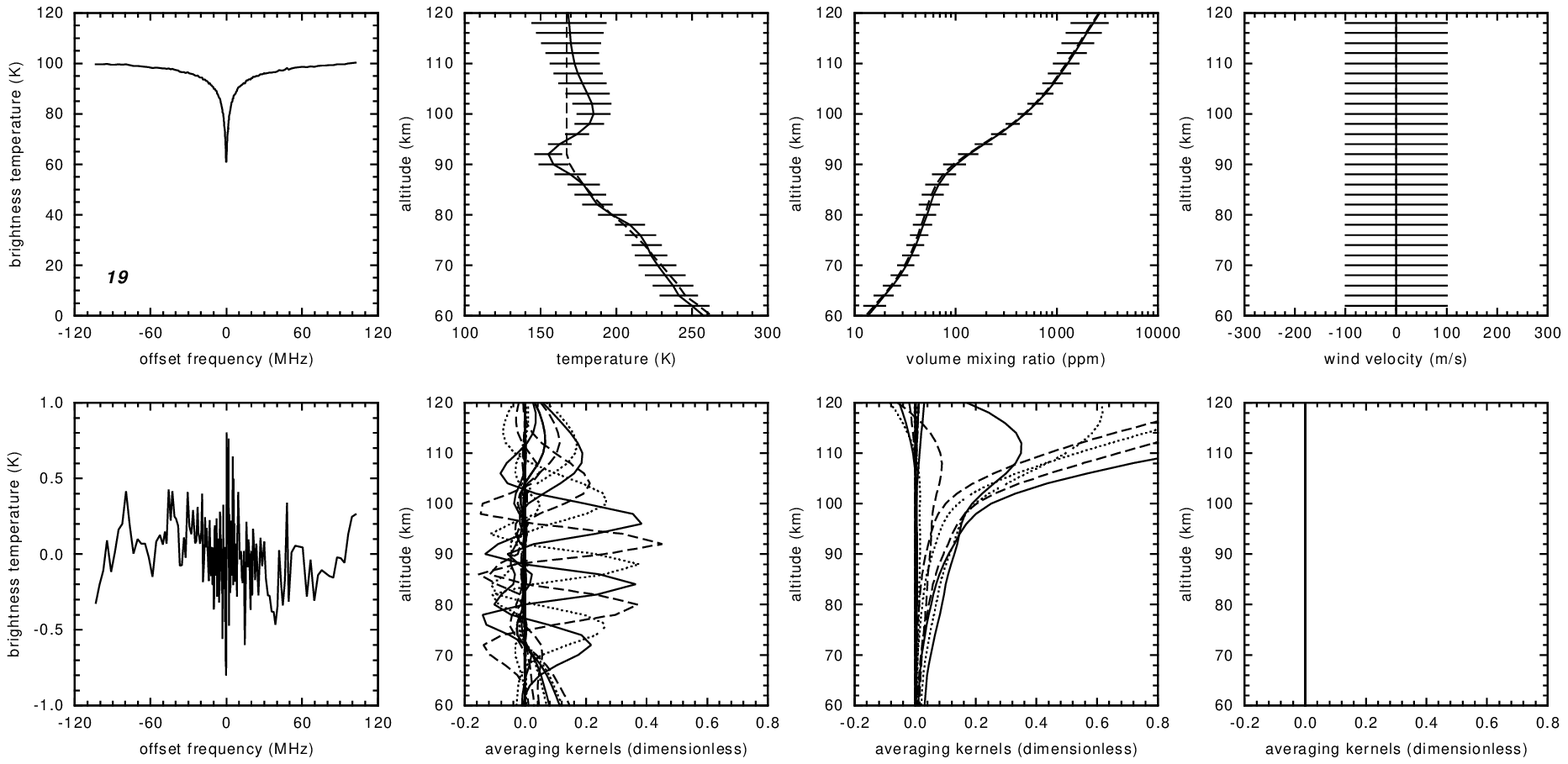,width=10.5cm} \psfig{file=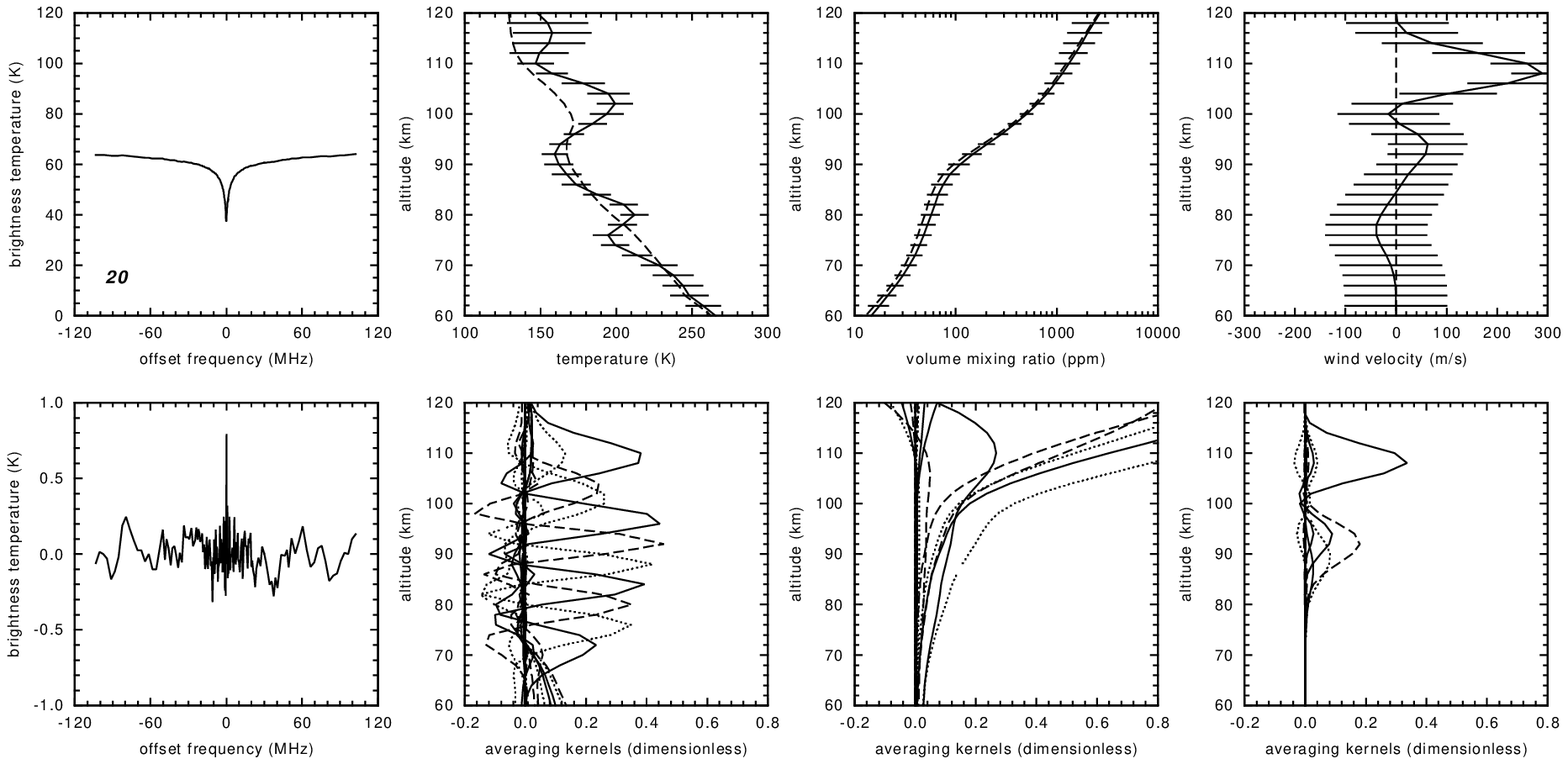,width=10.5cm}
\psfig{file=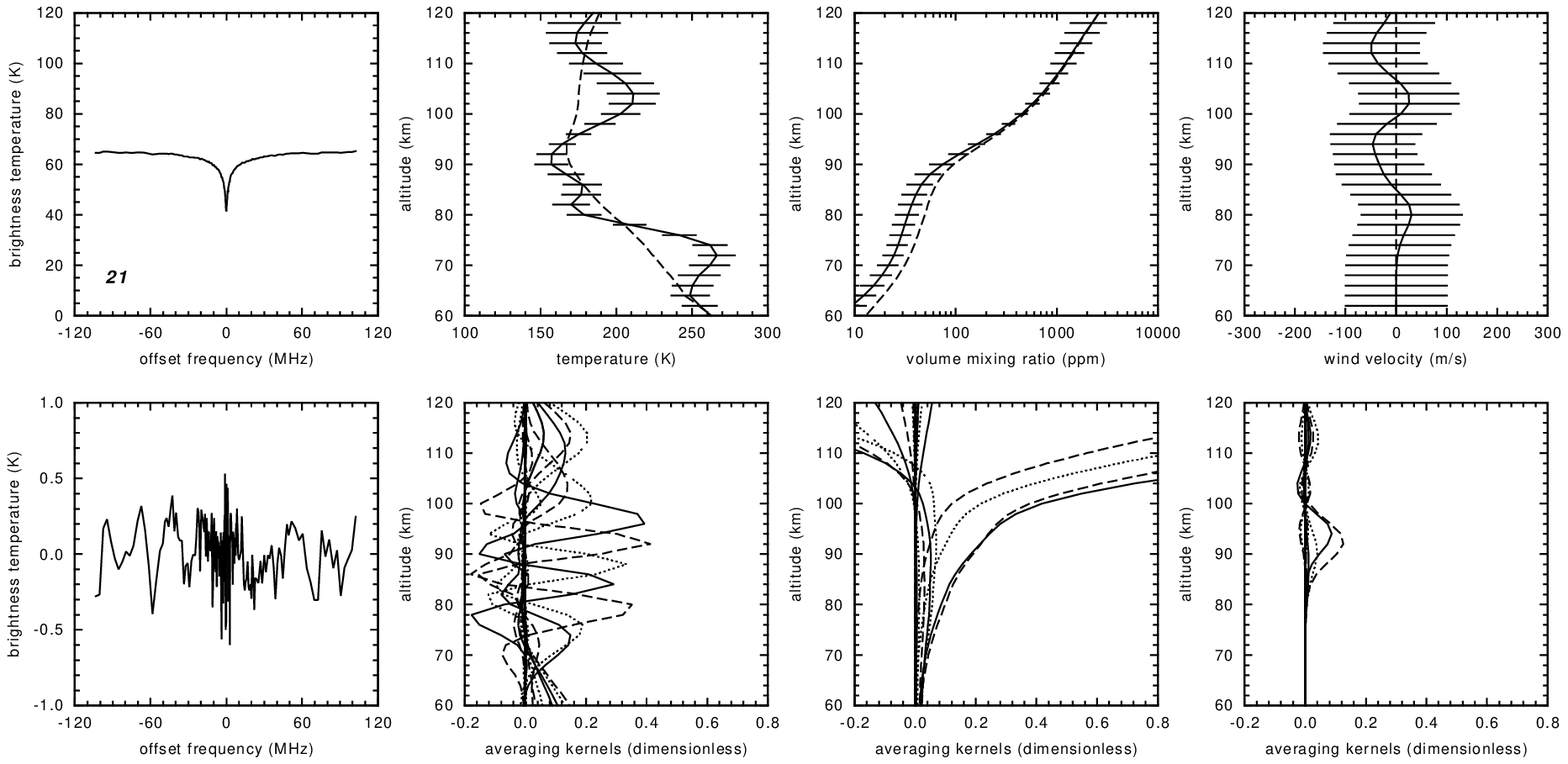,width=10.5cm}
\end{center}
\caption{}
\label{pag1}
\end{figure}

\begin{figure}
\begin{center}
\psfig{file=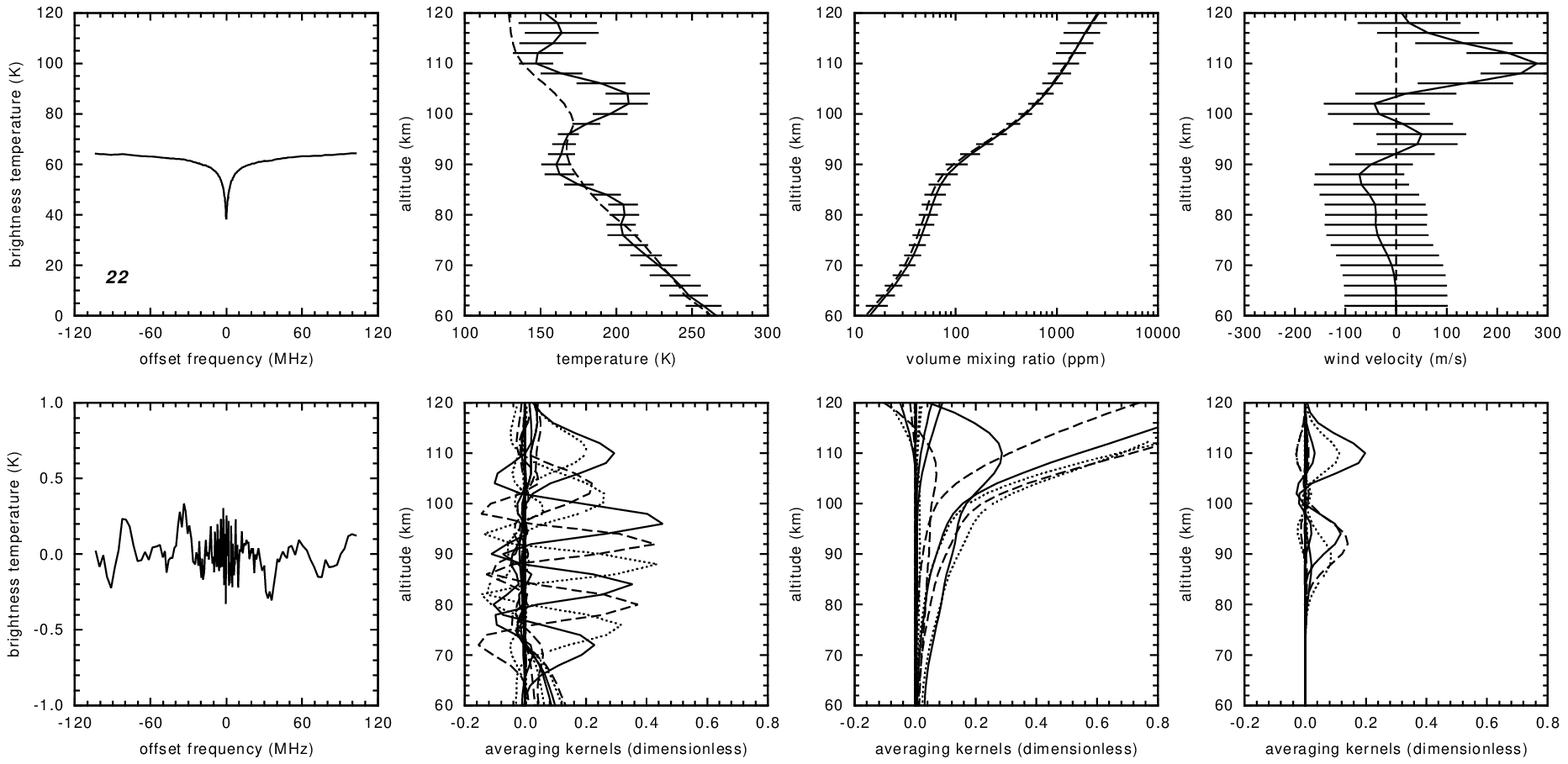,width=10.5cm} \psfig{file=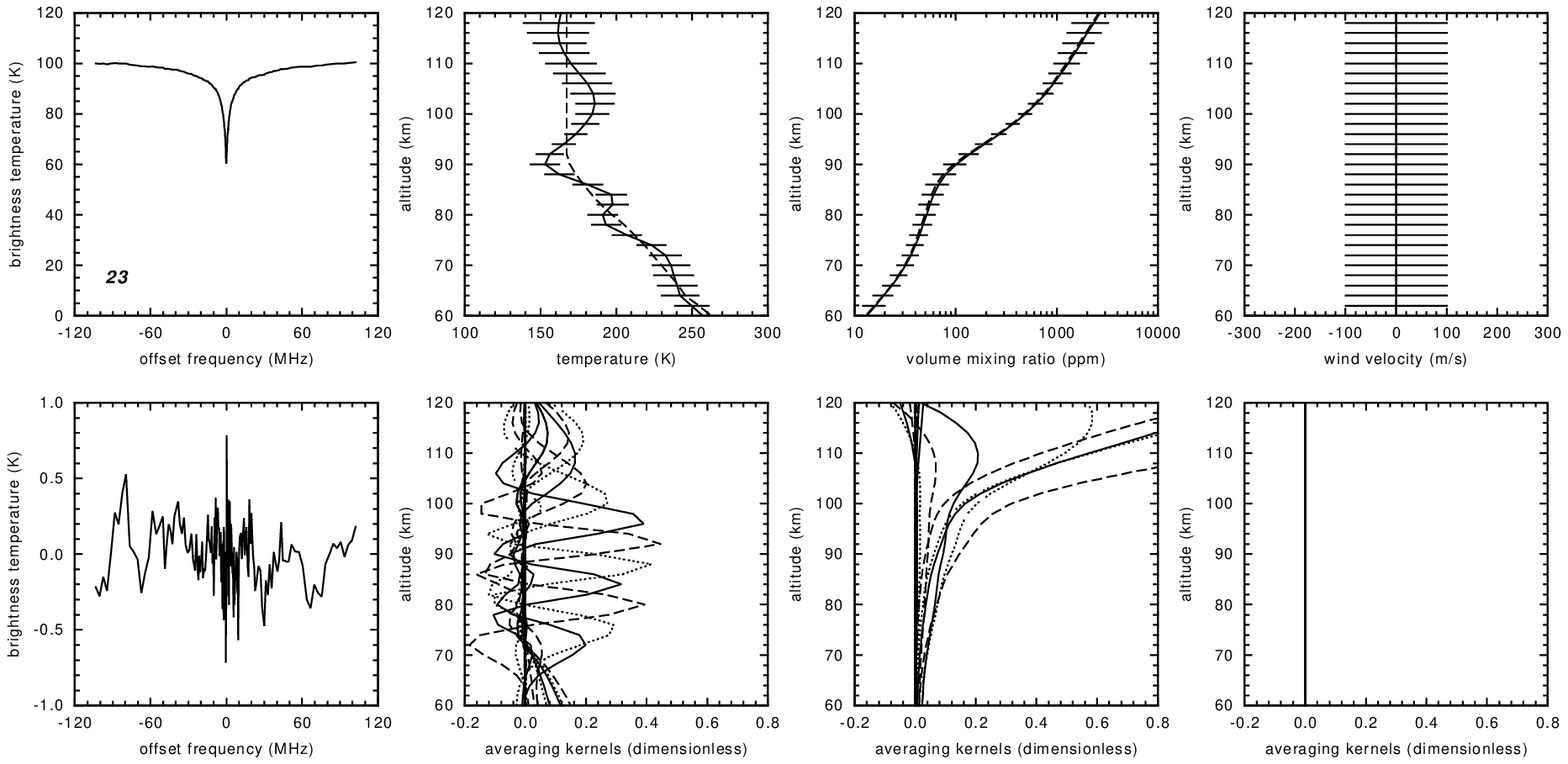,width=10.5cm}
\psfig{file=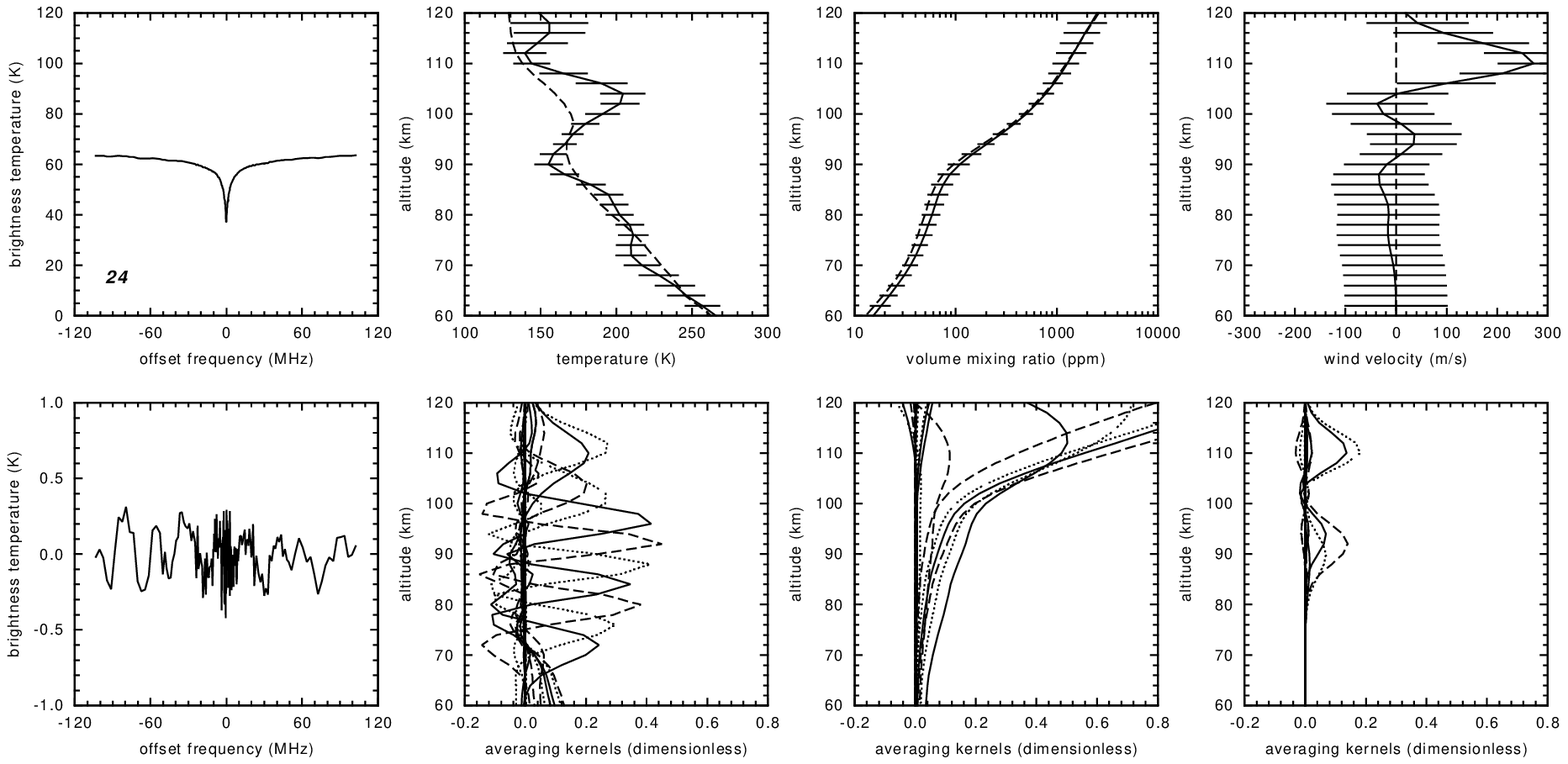,width=10.5cm}
\end{center}
\caption{}
\label{pag2}
\end{figure}

\begin{figure}
\begin{center}
\psfig{file=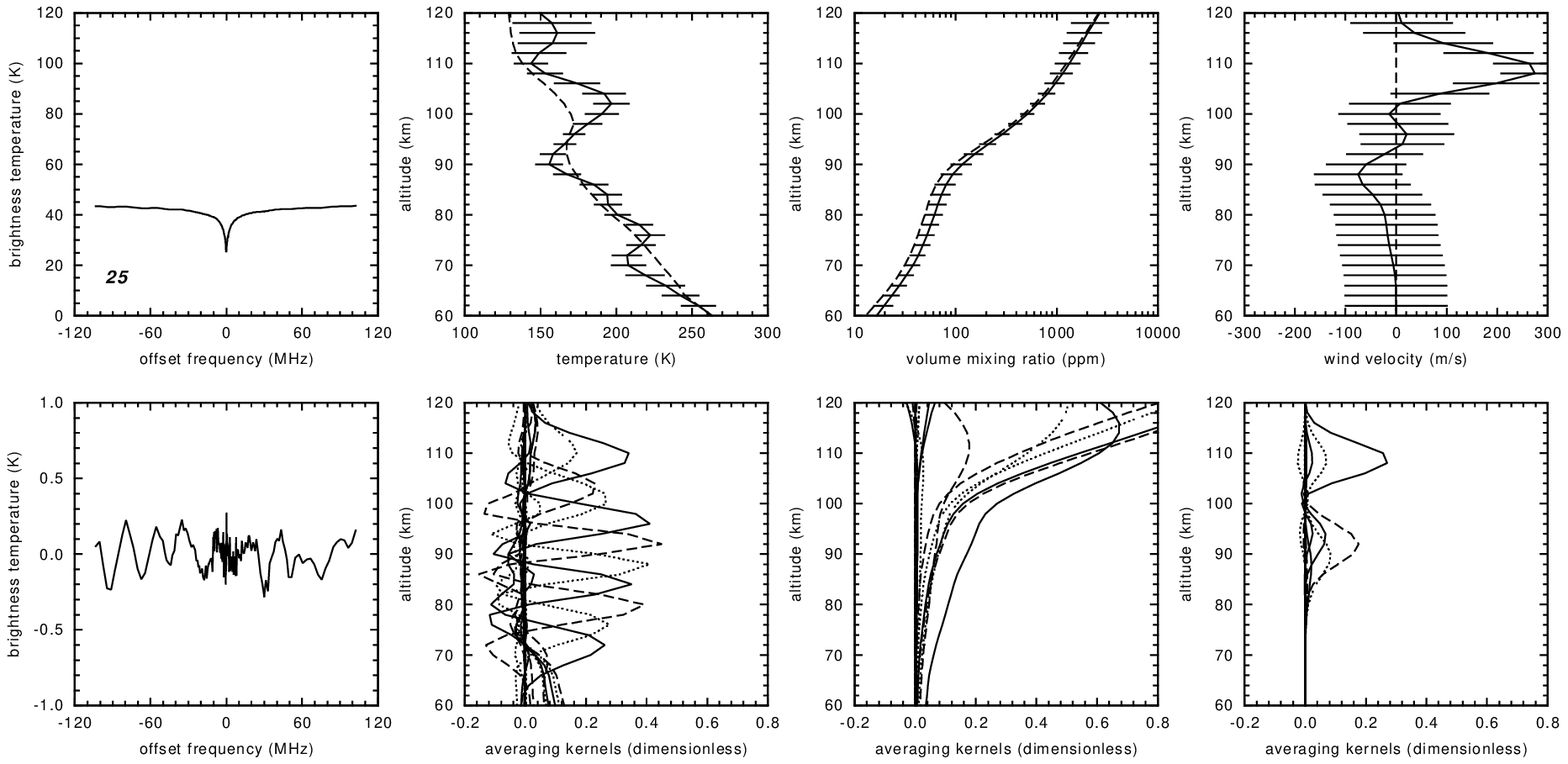,width=10.5cm} \psfig{file=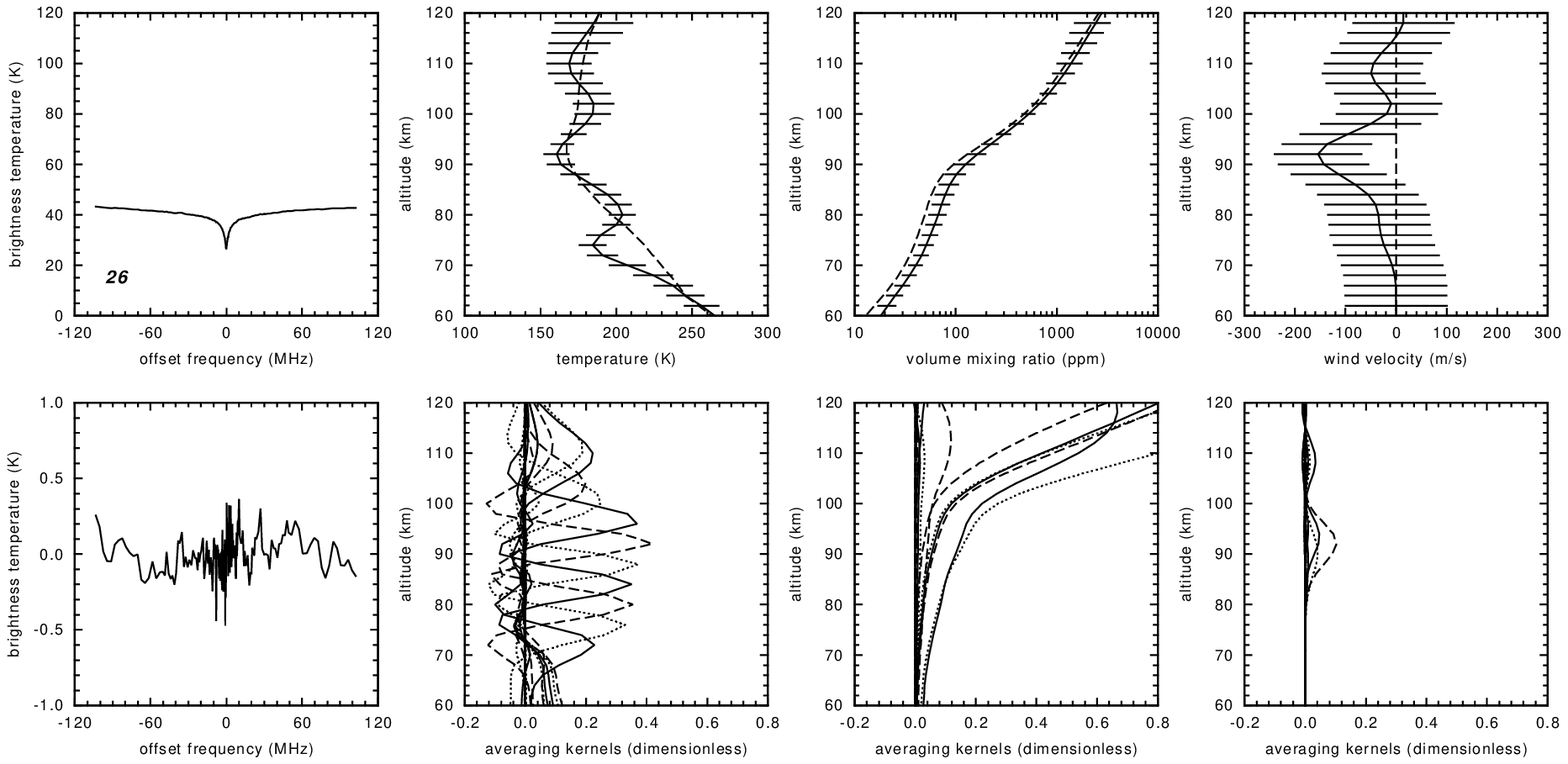,width=10.5cm}
\psfig{file=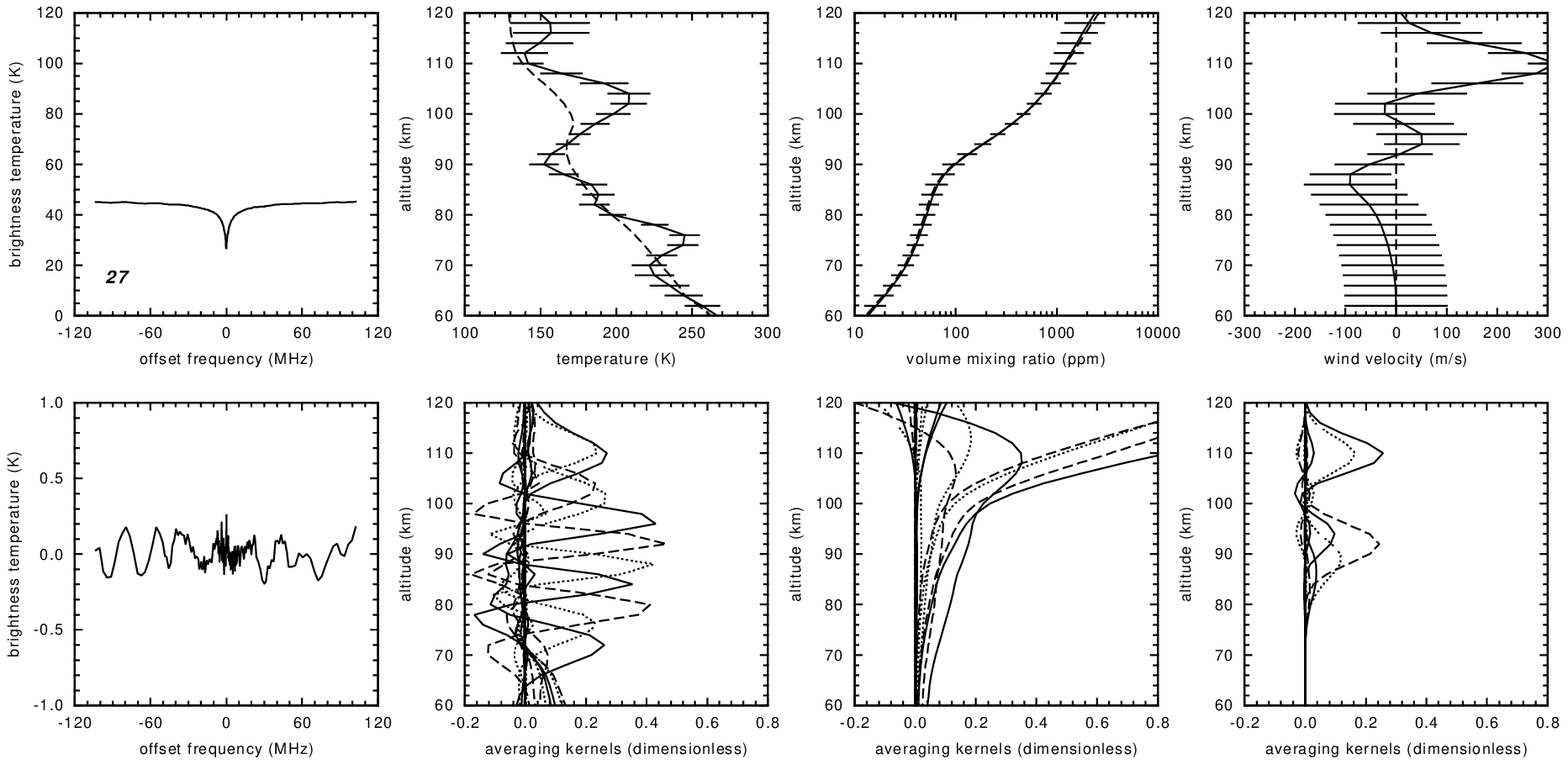,width=10.5cm}
\end{center}
\caption{}
\label{pag3}
\end{figure}

\begin{figure}
\begin{center}
\psfig{file=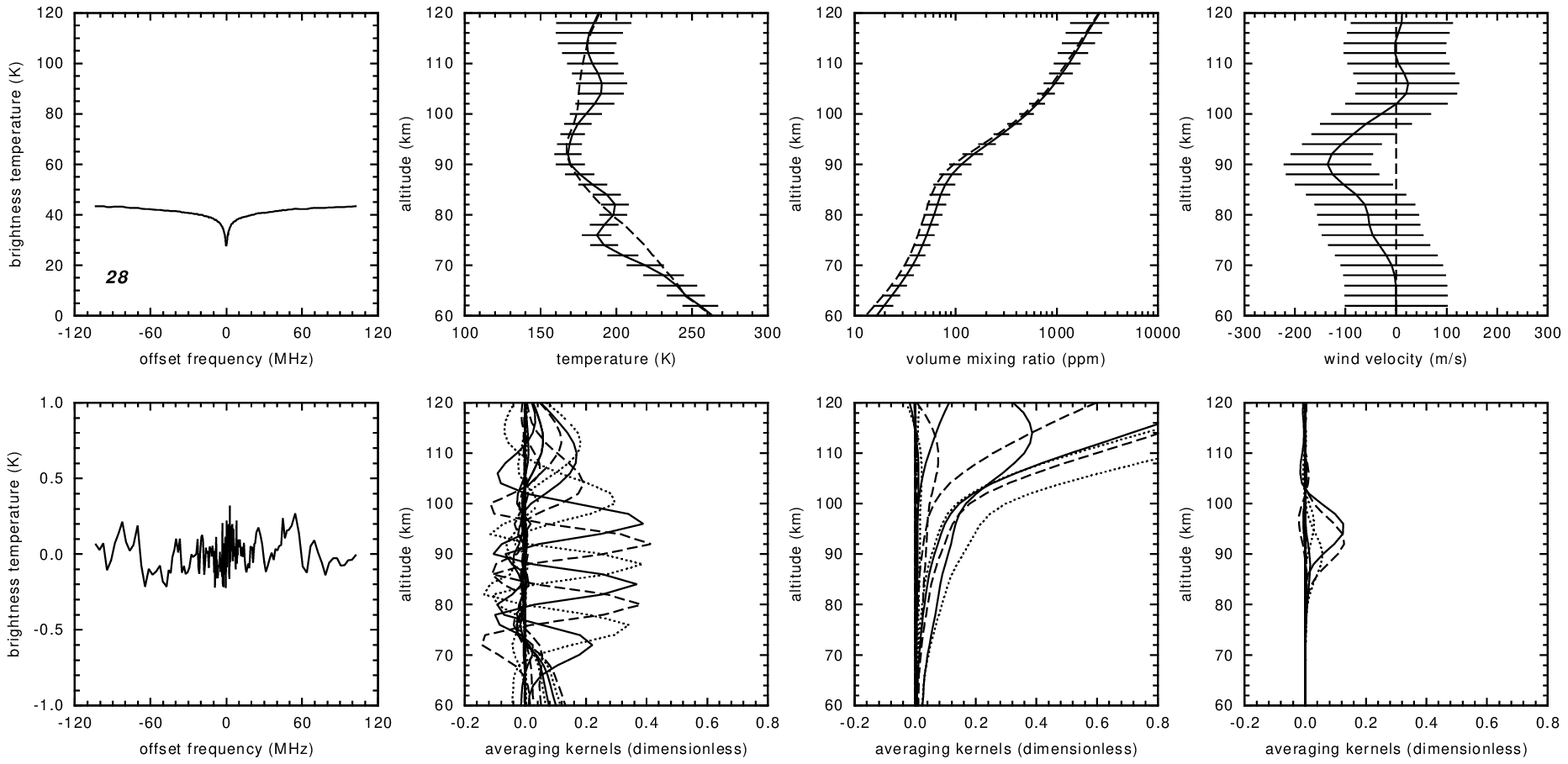,width=10.5cm} \psfig{file=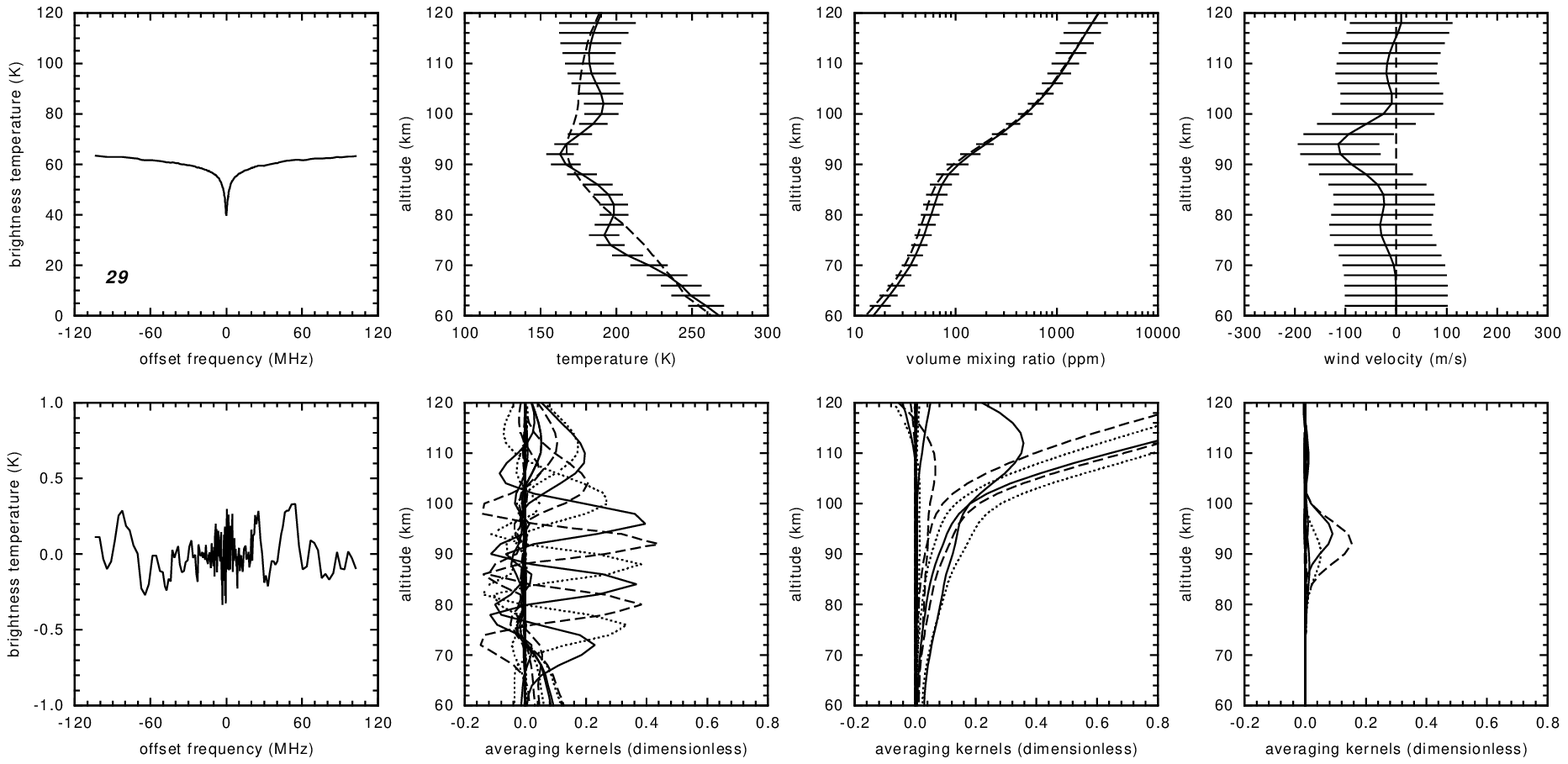,width=10.5cm}
\psfig{file=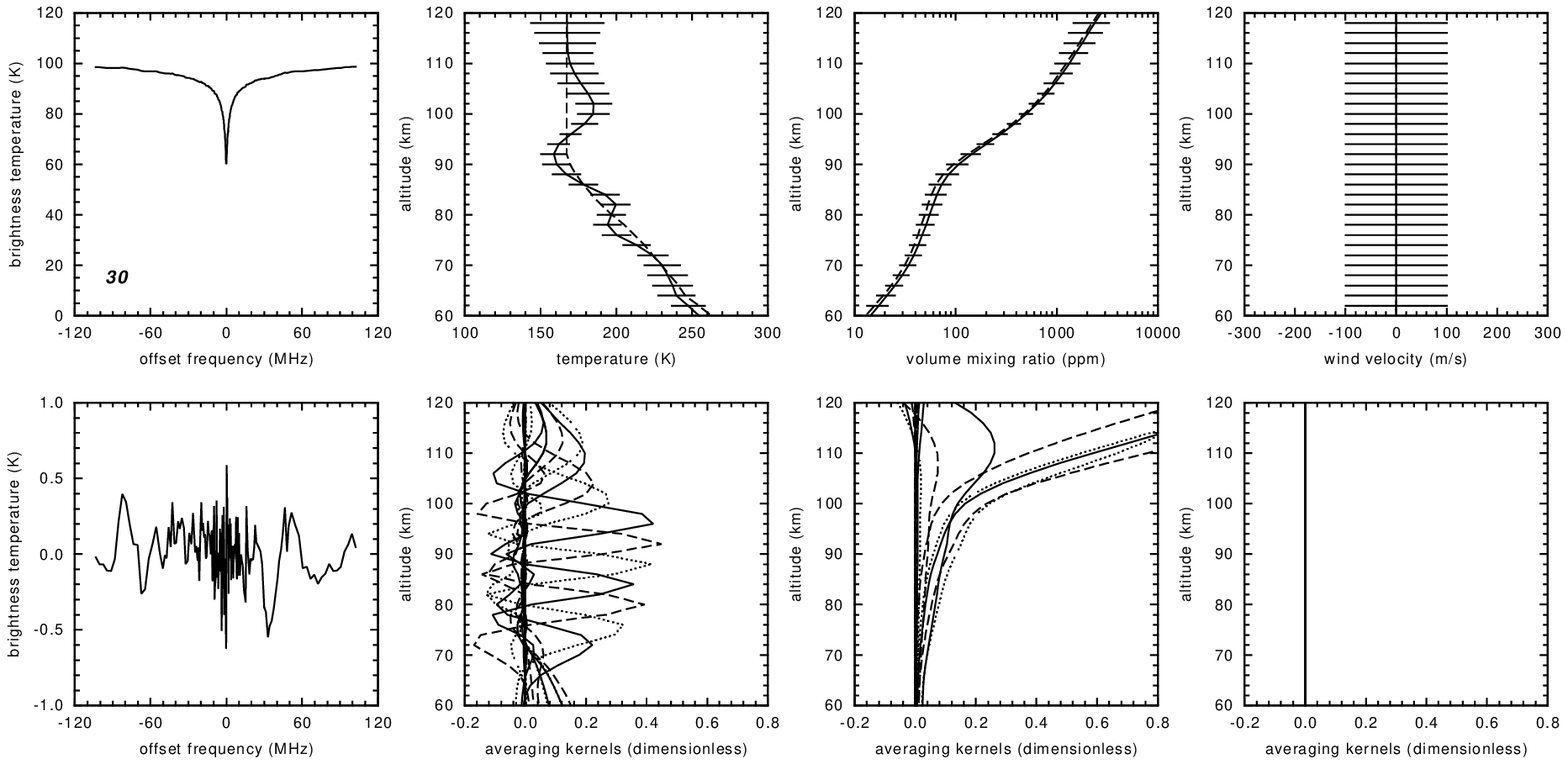,width=10.5cm}
\end{center}
\label{pag4}
\caption{}
\end{figure}

\begin{figure}
\begin{center}
\psfig{file=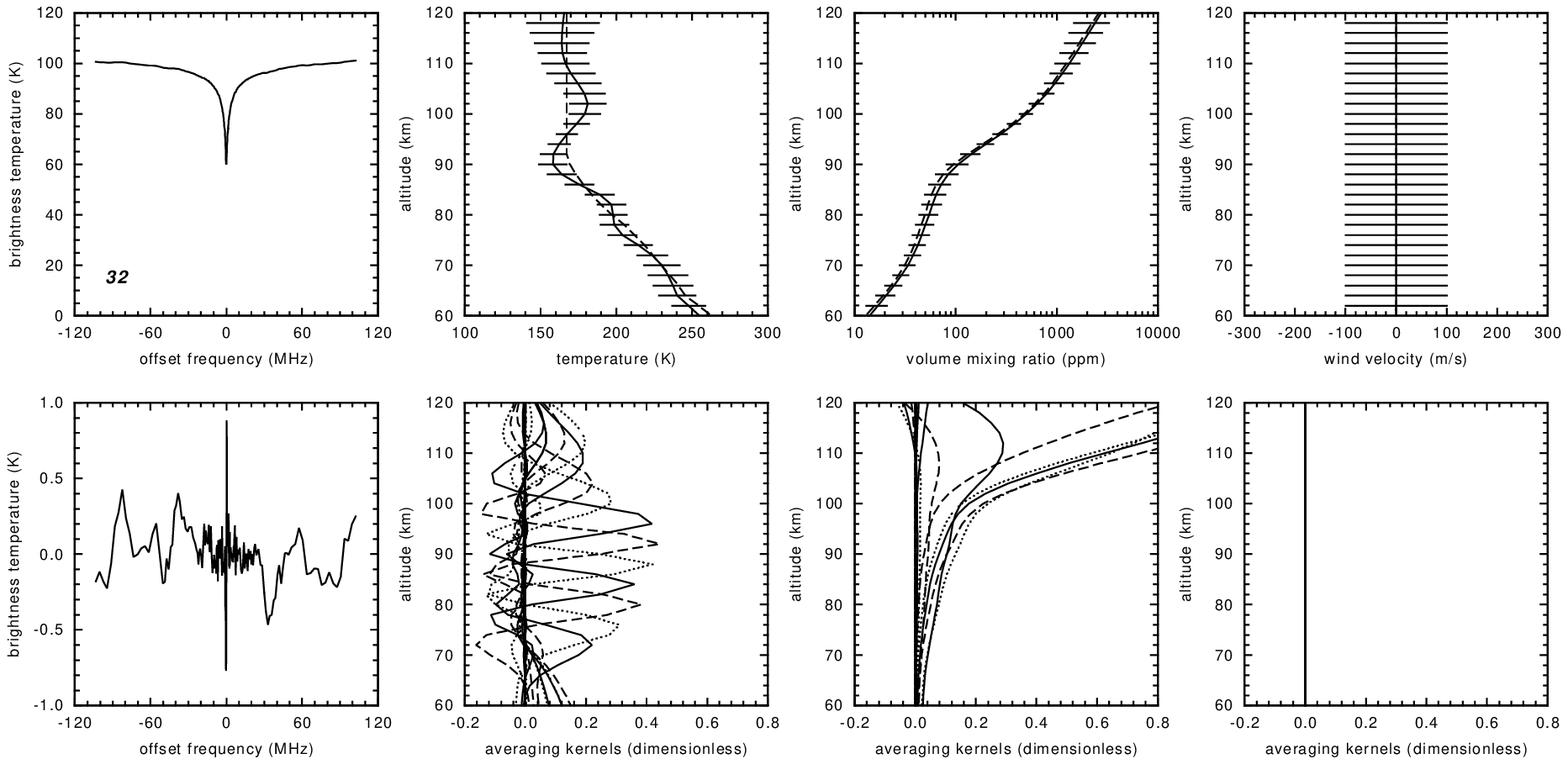,width=10.5cm} \psfig{file=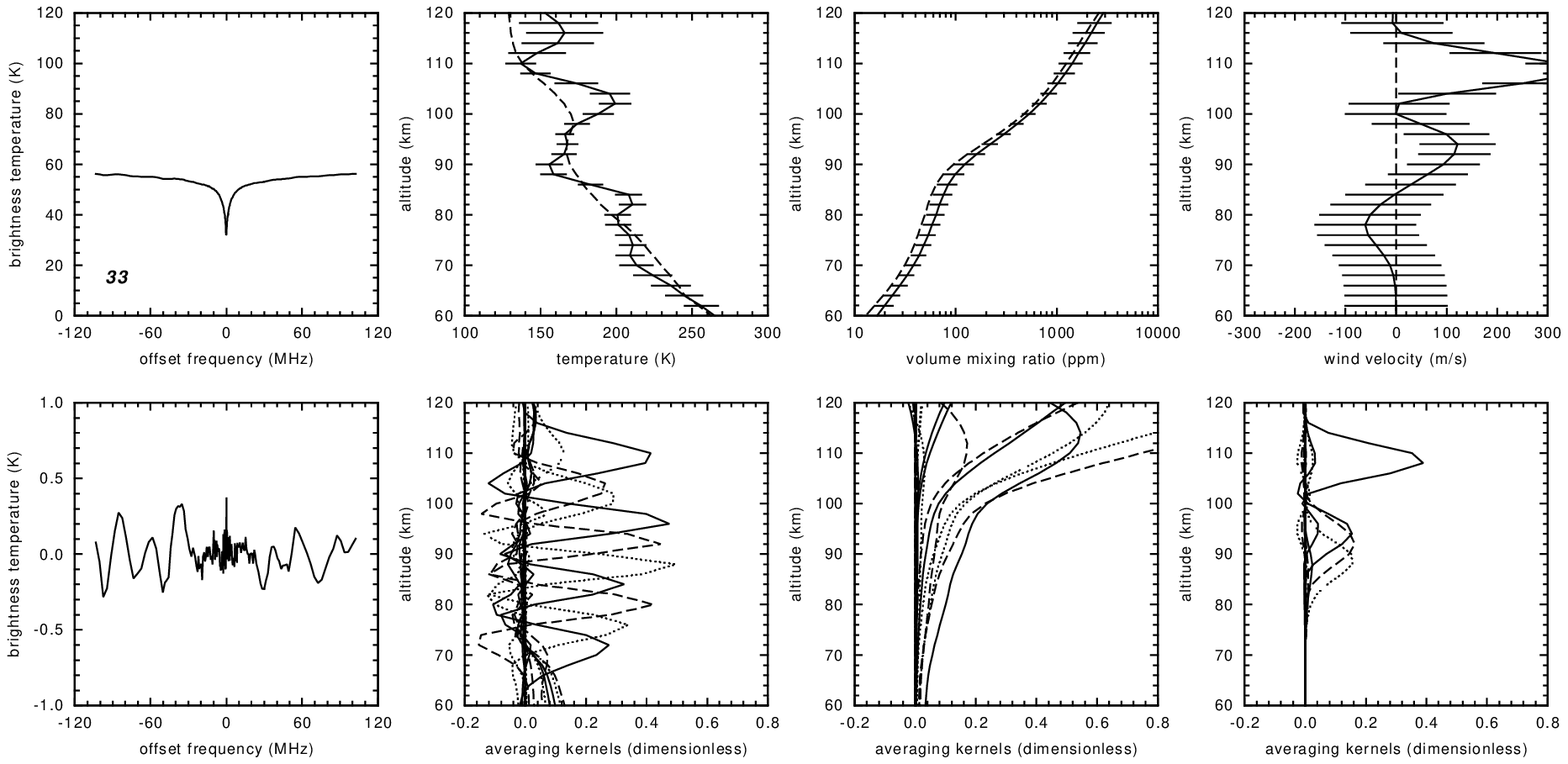,width=10.5cm}
\caption{}
\end{center}
\label{pag5}
\end{figure}

\begin{figure}
\begin{center}
\psfig{file=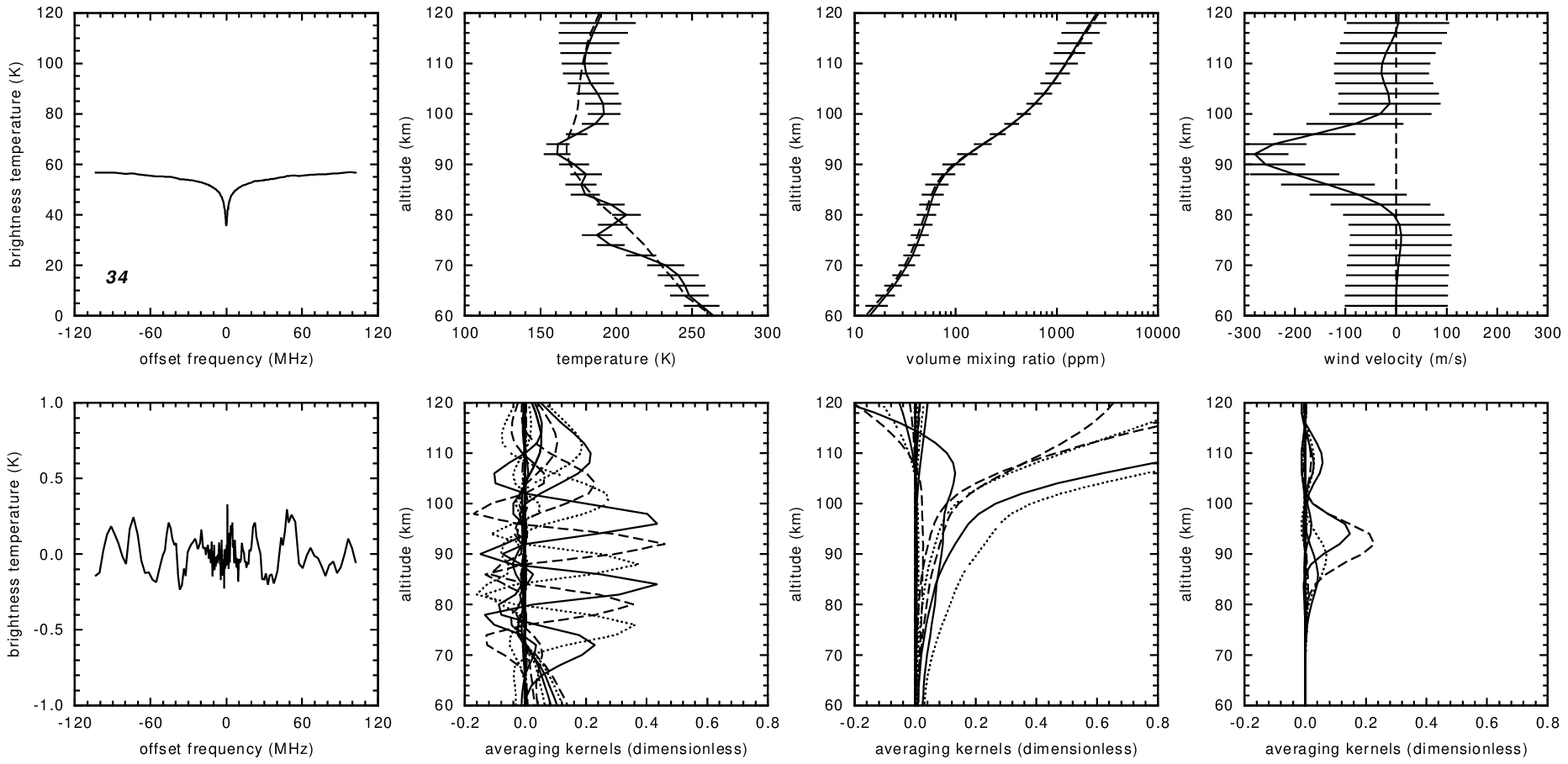,width=10.5cm} \psfig{file=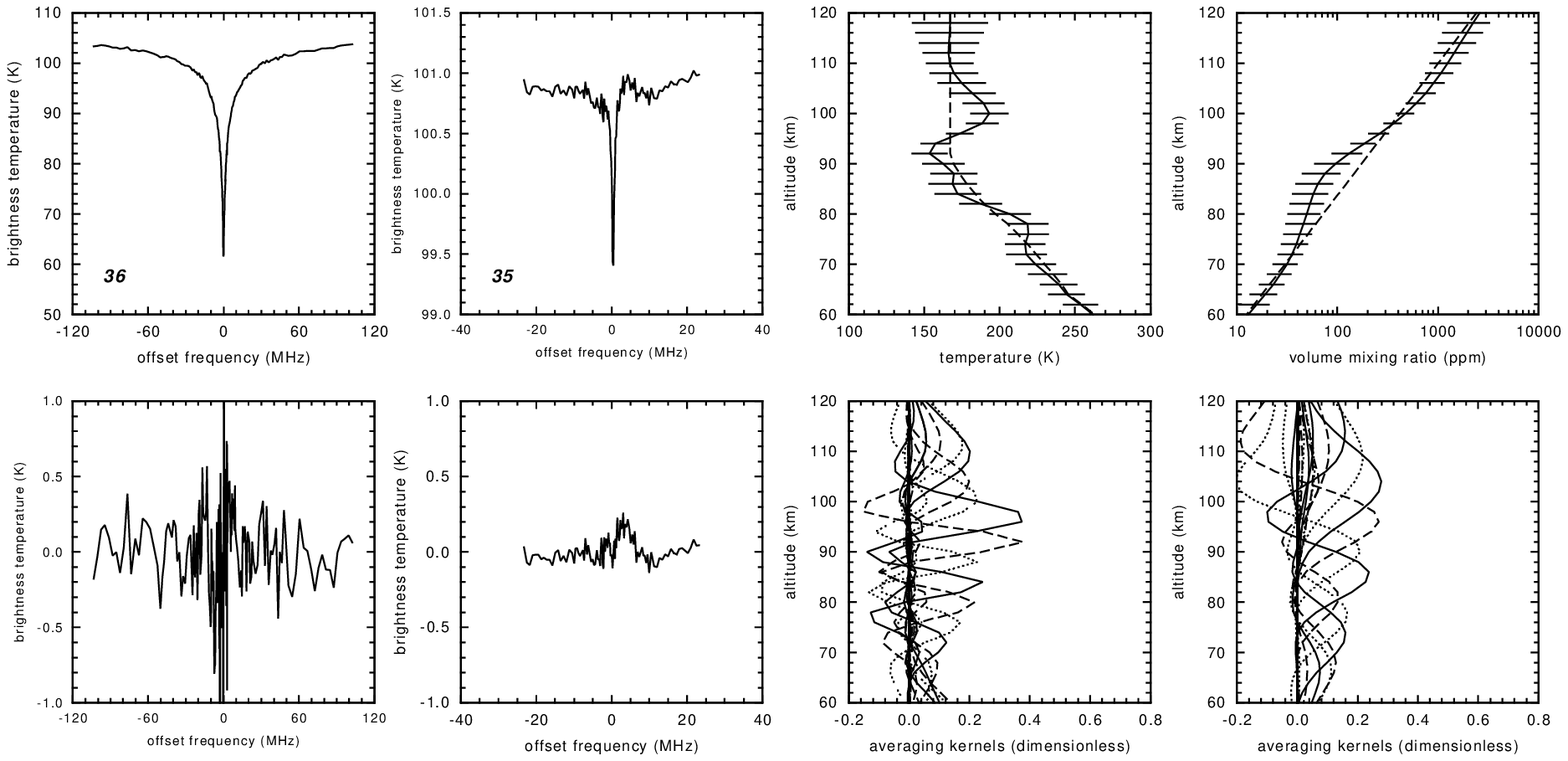,width=10.5cm}
\caption{}
\end{center}
\label{pag6}
\end{figure}

\begin{figure}
\begin{center}
\psfig{file=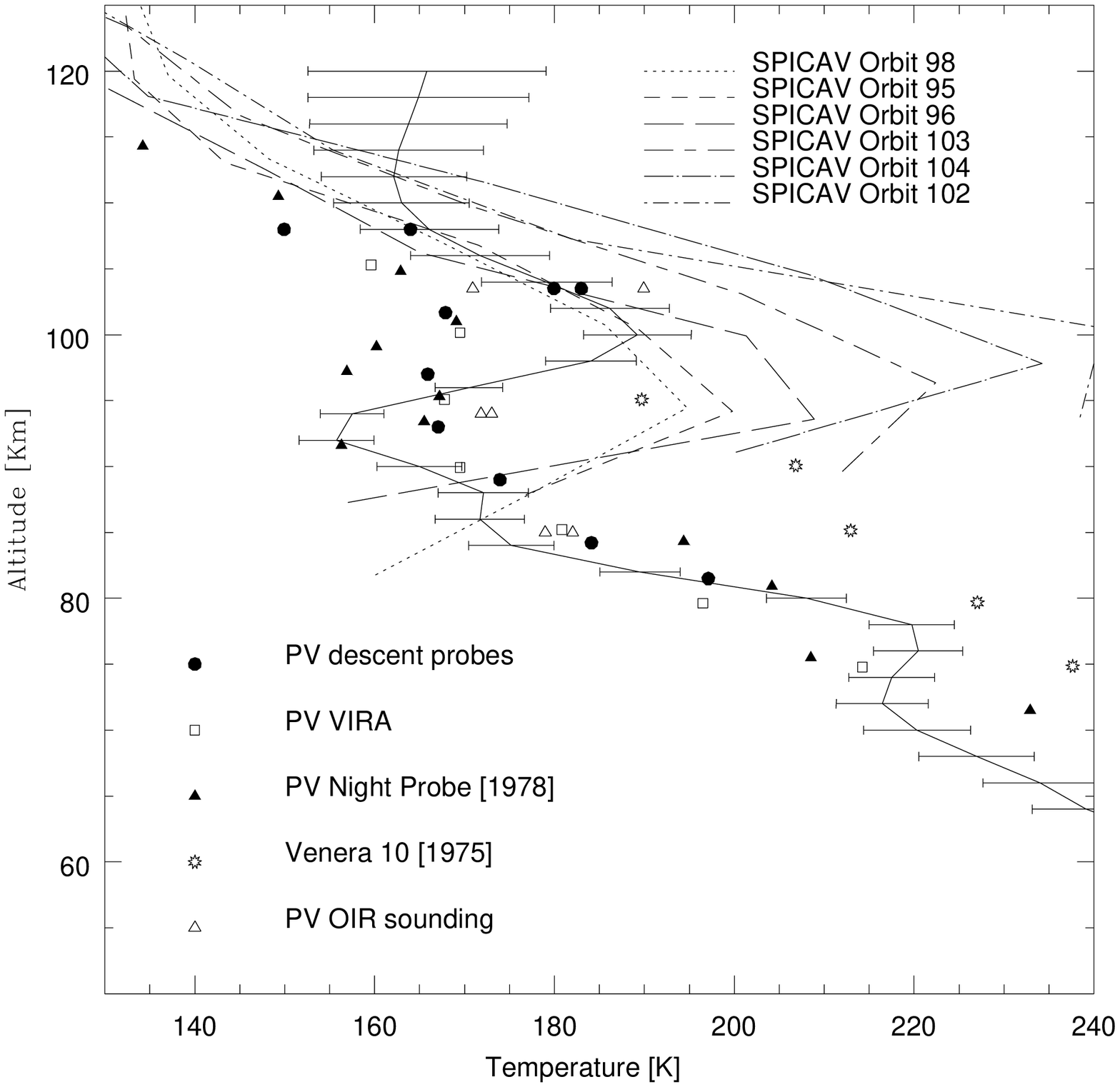,width=9cm}
\end{center}
\caption{} \label{compa}
\end{figure}

\begin{figure}[H]
\begin{center}
\psfig{file=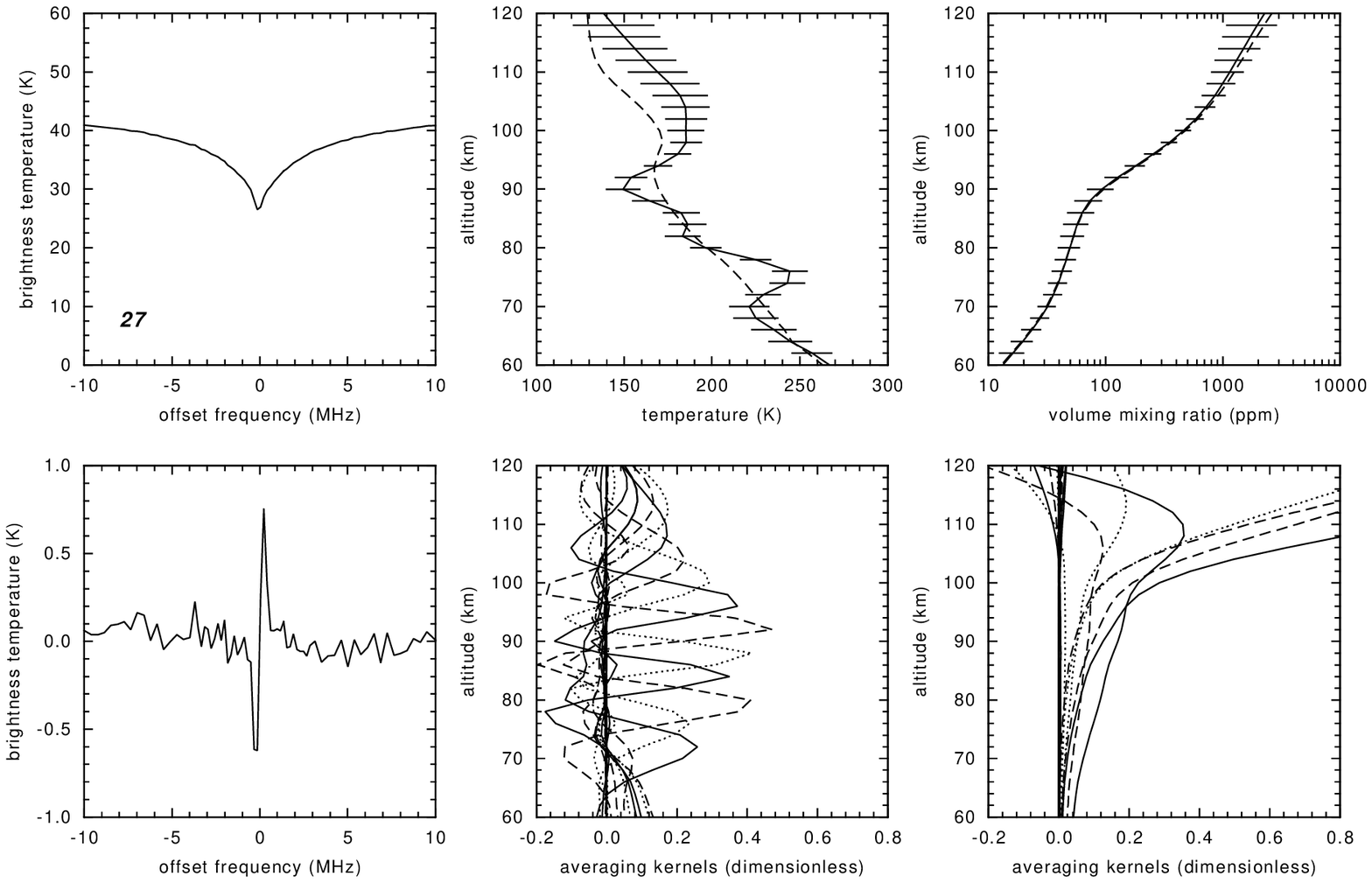,width=9cm}
\end{center}
\caption{} \label{wf3}
\end{figure}

\begin{figure}[H]
\begin{center}
\psfig{file=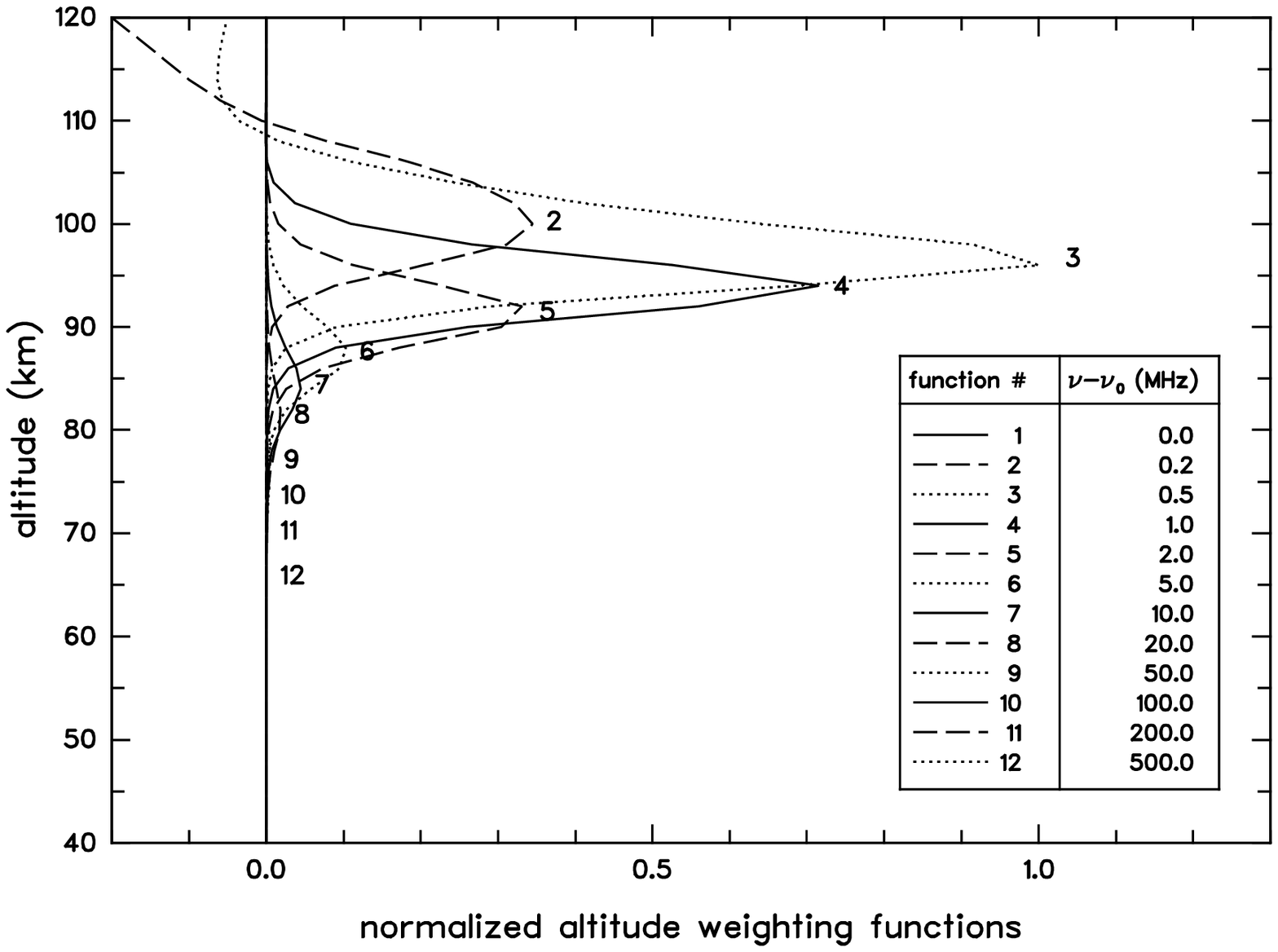,width=6cm}
\psfig{file=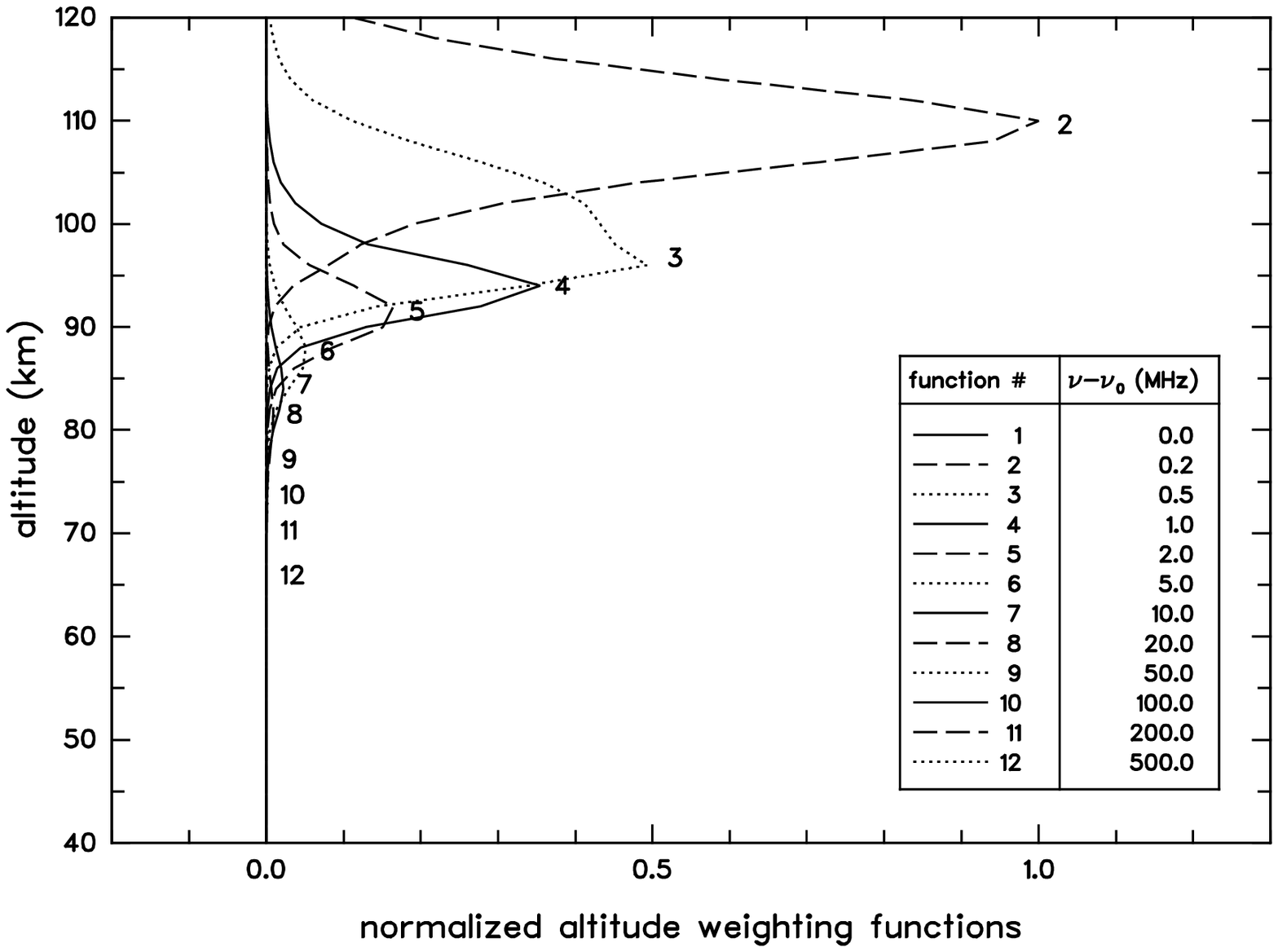,width=6cm}
\end{center}
\caption{} \label{wf4}
\end{figure}


\begin{figure}
\begin{center}
\psfig{file=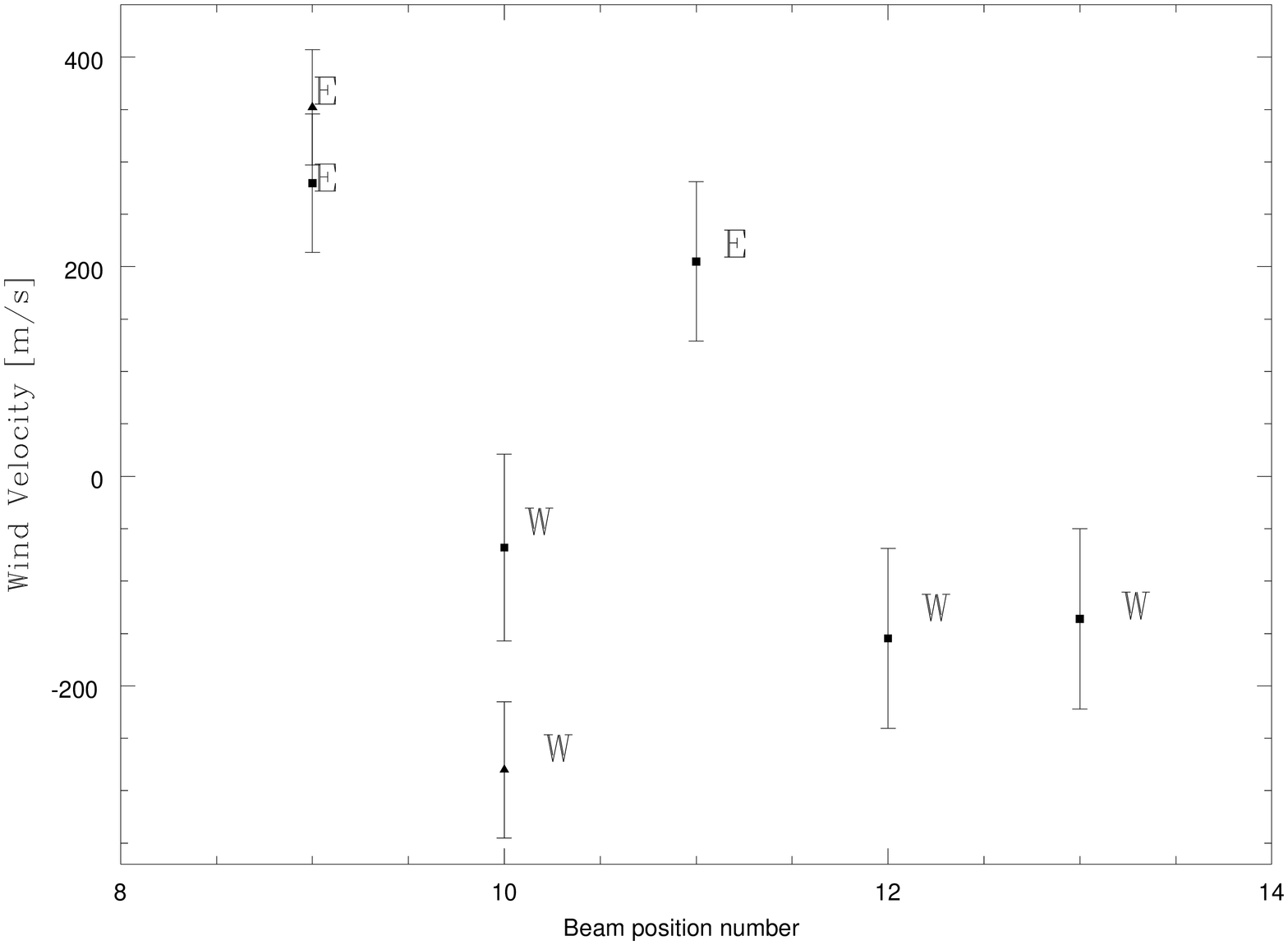,angle=-90,width=12cm}
\end{center}
\caption{}
\label{wind2}
\end{figure}

\begin{figure}
\begin{center}
\psfig{file=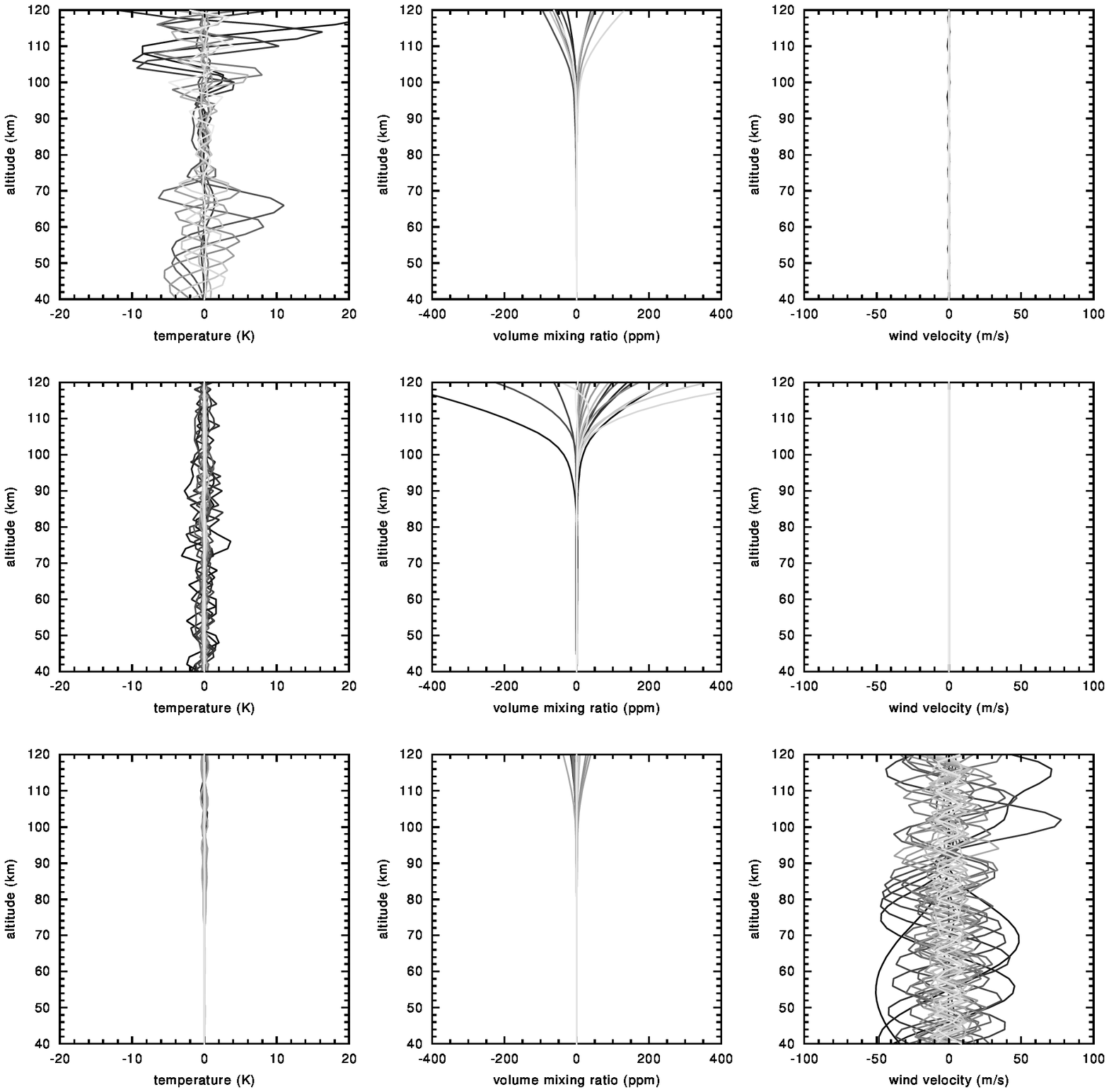,width=11cm}
\end{center}
\caption{} \label{cerr1}
\end{figure}

\begin{figure}
\begin{center}
\psfig{file=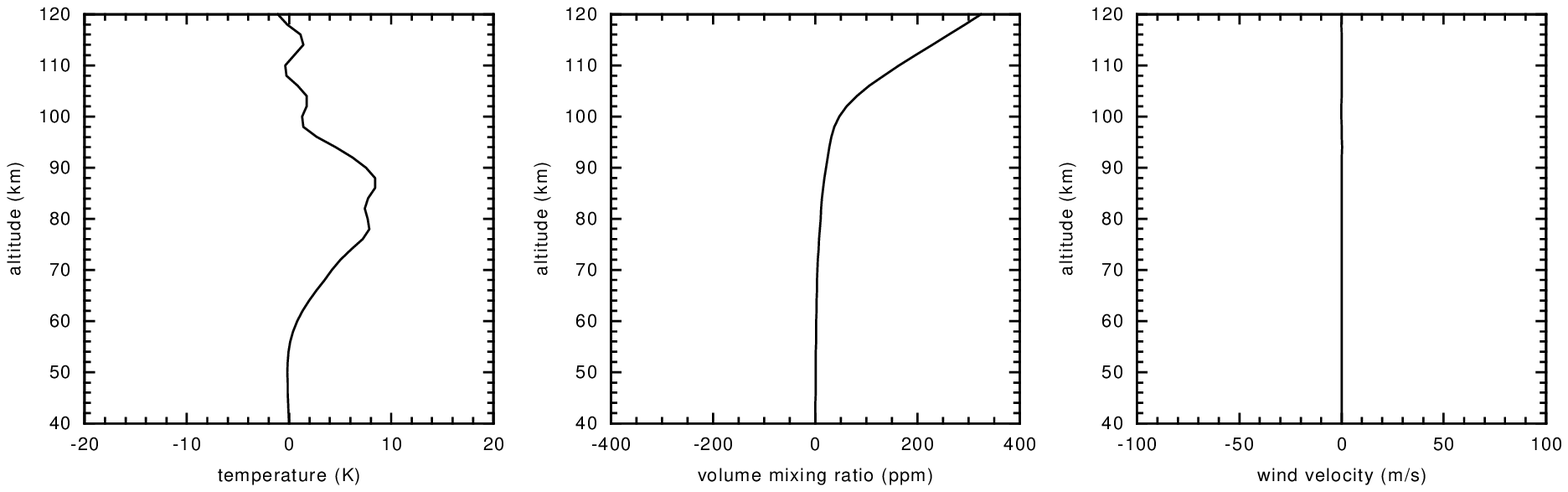,width=11cm}
\end{center}
\caption{}
\label{cerr2}
\end{figure}

\clearpage

\begin{figure}[H]
\begin{center}
\psfig{file=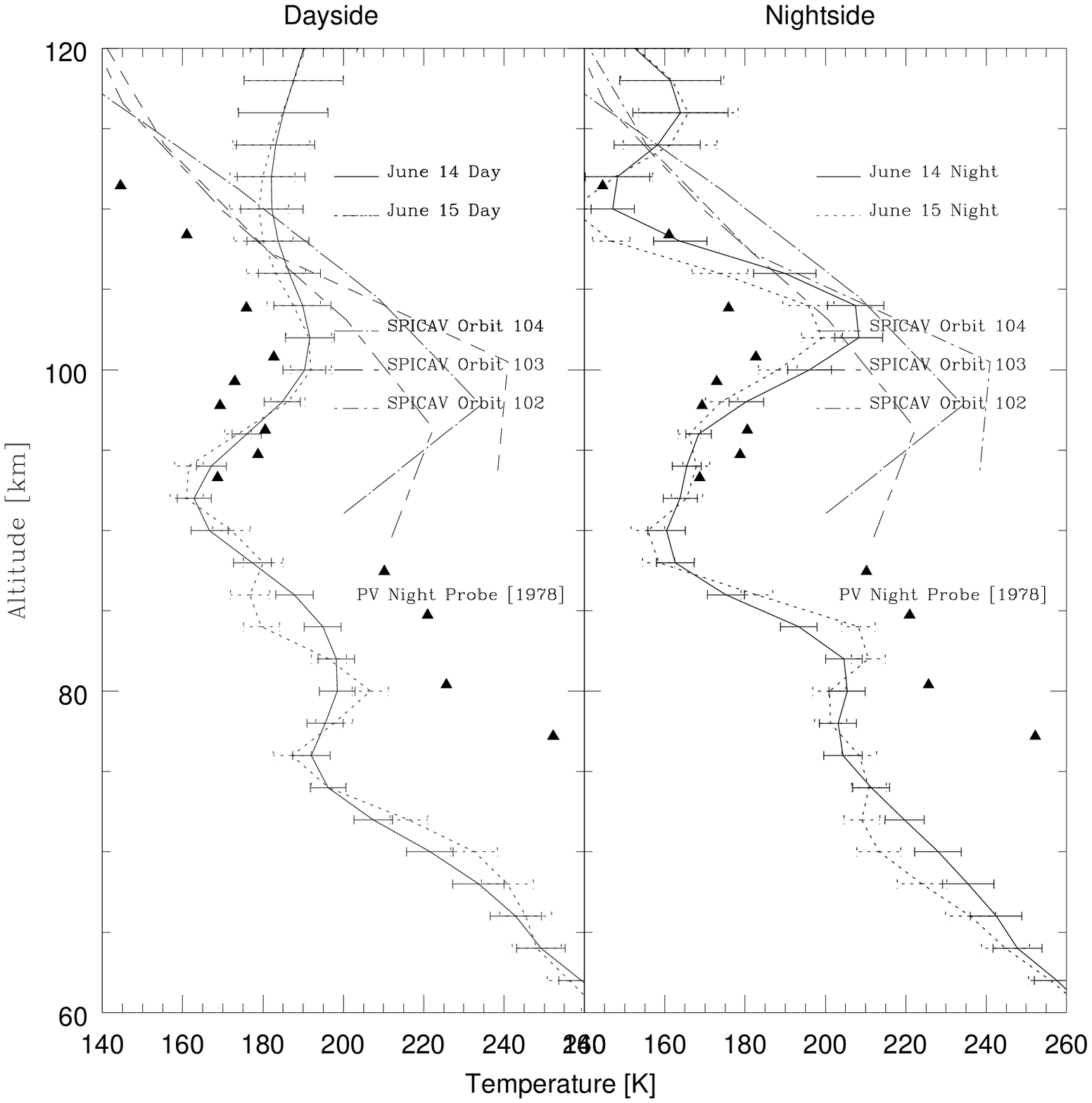,width=9cm}
\end{center}
\caption{}
\label{example}
\end{figure}

\end{linenumbers}
\end{document}